%% file: 1-main-acm.tex
\documentclass[acmsmall]{acmart}

\AtBeginDocument{%
  }

\usepackage{tabularx}
\usepackage{graphicx}
\usepackage{subcaption}
\usepackage{multirow}
\usepackage{enumitem}
\usepackage{appendix}

\newcommand{\revise}[1]{\textcolor{black}{#1}}
\newcommand{\minor}[1]{\textcolor{black}{#1}}
\newcommand{\camera}[1]{\textcolor{black}{#1}}

\begin{document}



\title[Script-Strategy Aligned Generation]{Script-Strategy Aligned Generation: Aligning LLMs with Expert-Crafted Dialogue Scripts and Therapeutic Strategies for Psychotherapy}




\author{Xin Sun}
\email{x.sun2@uva.nl}
\affiliation{%
  \institution{University of Amsterdam}
  \city{Amsterdam}
  \state{}
  \country{The Netherlands}
  \institution{and National Institute of Informatics (NII)}
  \city{Tokyo}
  \state{}
  \country{Japan}
}

\author{Jan de Wit}
\affiliation{%
  \institution{Tilburg University}
  \city{Tilburg}
  \state{}
  \country{The Netherlands}
}

\author{Zhuying Li}
\affiliation{%
  \institution{Southeast University}
  \country{China}
}

\author{Jiahuan Pei}
\affiliation{%
  \institution{Vrije Universiteit Amsterdam}
  \city{Amsterdam}
  \country{The Netherlands}
}

\author{Abdallah El Ali}
\affiliation{%
  \institution{Centrum Wiskunde \& Informatica (CWI) and Utrecht University}
  \city{Amsterdam}
  \country{The Netherlands}
}

\author{Jos A. Bosch}
\affiliation{%
  \institution{University of Amsterdam}
  \city{Amsterdam}
  \state{}
  \country{The Netherlands}
}










\begin{abstract}
\camera{Chatbots or conversational agents (CAs) are increasingly used to improve access to digital psychotherapy. Many current systems rely on rigid, rule-based designs, heavily dependent on expert-crafted dialogue scripts for guiding therapeutic conversations. 
Although advances in large language models (LLMs) offer potential for more flexible interactions, their lack of controllability and explanability poses challenges in high-stakes contexts like psychotherapy.
To address this, we conducted two studies in this work to explore how aligning LLMs with expert-crafted scripts can enhance psychotherapeutic chatbot performance. 
In Study 1 (N=43), an online experiment with a within-subjects design, we compared rule-based, pure LLM, and LLMs aligned with expert-crafted scripts via fine-tuning and prompting. 
Results showed that aligned LLMs significantly outperformed the other types of chatbots in empathy, dialogue relevance, and adherence to therapeutic principles.
Building on findings, we proposed ``Script-Strategy Aligned Generation (SSAG)'', a more flexible alignment approach that reduces reliance on fully scripted content while maintaining LLMs' therapeutic adherence and controllability. 
In a 10-day field Study 2 (N=21), SSAG achieved comparable therapeutic effectiveness to full-scripted LLMs while requiring less than 40\% of expert-crafted dialogue content. 
Beyond these results, this work advances LLM applications in psychotherapy by providing a controllable and scalable solution, reducing reliance on expert effort. 
By enabling domain experts to align LLMs through high-level strategies rather than full scripts, SSAG supports more efficient co-development and expands access to a broader context of psychotherapy.}
\end{abstract}

\begin{teaserfigure}
\centering
\includegraphics[width=0.999\textwidth]{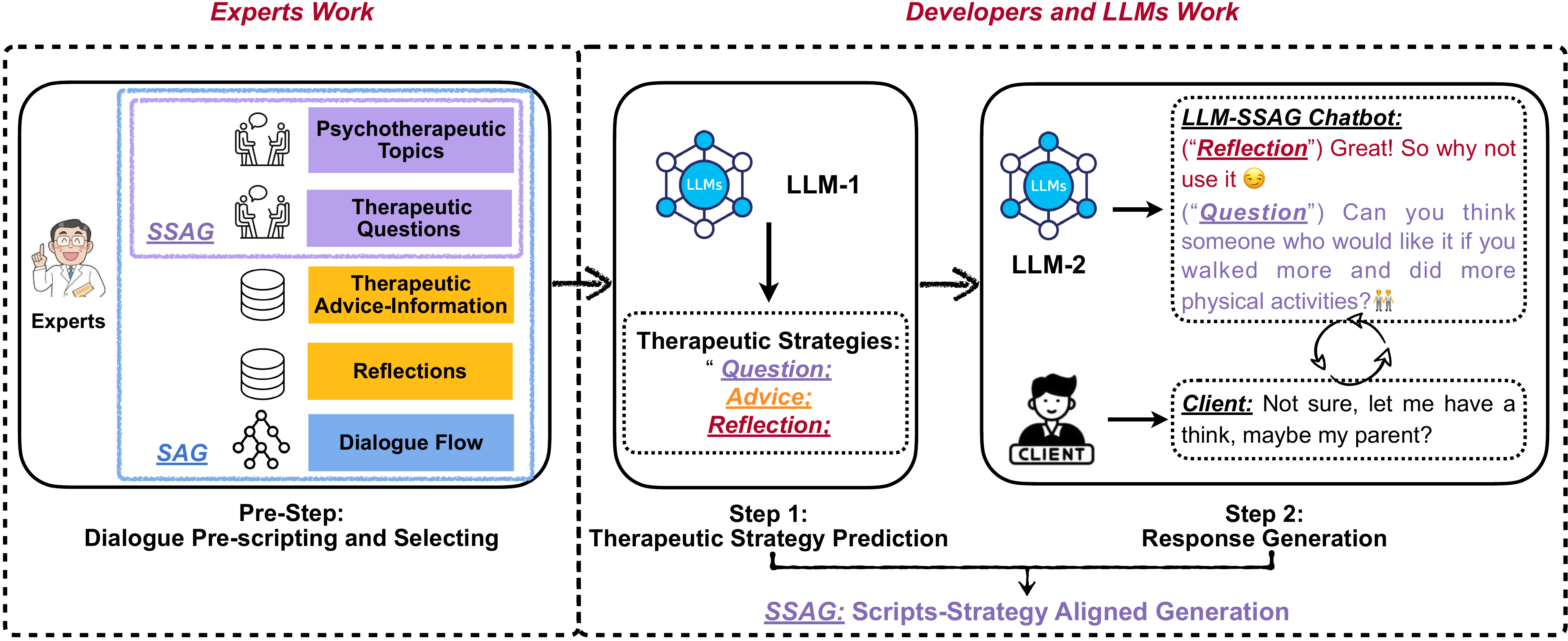}
\caption{
\revise{Visualization of SSAG and its comparison with SAG. In SSAG, experts provide core therapeutic content, while LLMs first predict therapeutic strategy and then generate responses accordingly. 
In comparison, SAG relies on fully expert-crafted dialogue scripts.}
}
\label{fig:ssag}
\end{teaserfigure}






\received{October 2024}
\received[revised]{April 2025}
\received[accepted]{August 2025}

\maketitle


\input{2-Introduction}

\input{3-Related}

\input{4-Dataset}

\input{5-Study-1-Methods}

\input{5-Study-1-Results}
\input{6-Study-2-Methods}
\input{6-Study-2-Results}

\input{7-Discussion}

\input{8-Conclusion}

\section*{Acknowledgment}
This work is funded by the European Commission in the Horizon H2020 scheme, awarded to Jos A. Bosch (TIMELY Grant agreement ID: 101017424). 
All content represents the opinion of the authors, which is not necessarily shared or endorsed by their respective employers and/or sponsors.



\newpage
\clearpage

\bibliographystyle{ACM-Reference-Format}
\bibliography{1-main-acm}


\input{appendix_dialogues}
\input{appendix_prompts}

\input{appendix_implement}

\end{document}

%% file: 2-Introduction.tex
\section{Introduction}

\begin{figure}[!ht]
\centering
\includegraphics[width=0.98\textwidth]{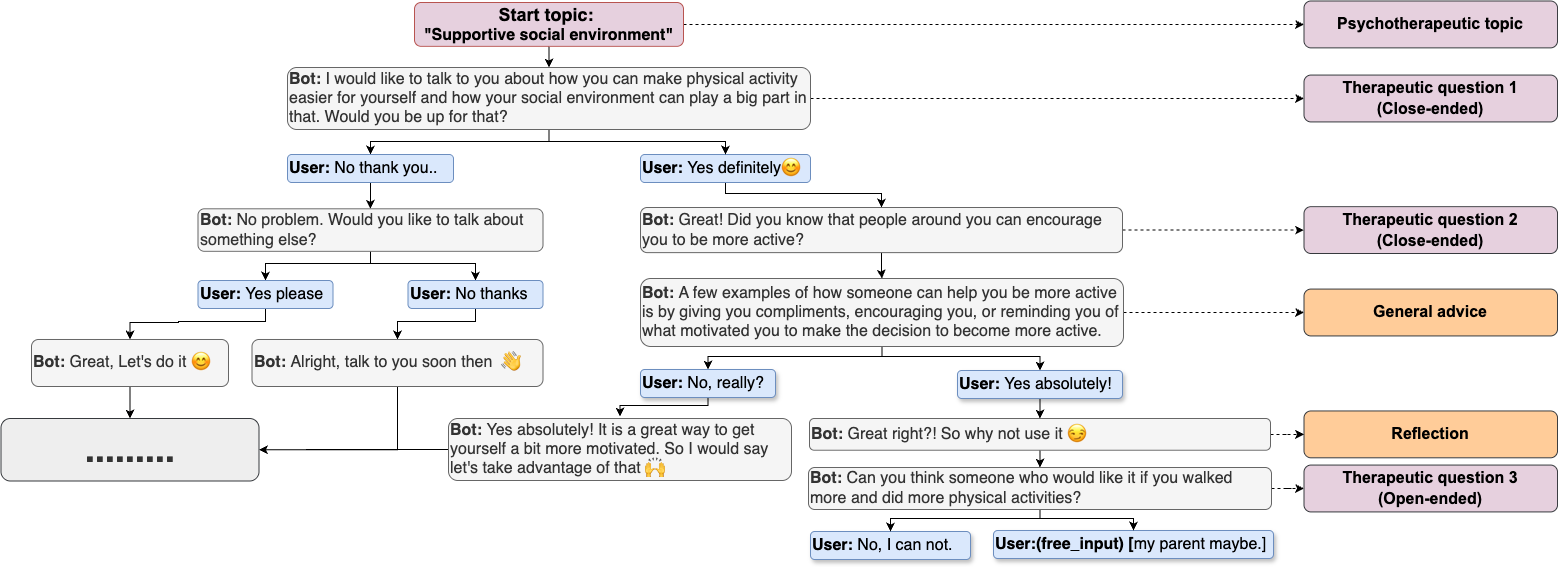}
\vspace{-1.6mm}
\caption{A dialogue example: parts of tree-structured dialogue scripts pre-crafted by experts under the psychotherapeutic topic ``Supportive social environment'' for behavioral intervention.}
\vspace{-2mm}
\label{fig:dataset_example}
\end{figure}

Chatbots, or conversational agents (CAs), are increasingly used in psychotherapy for behavioral interventions. These agents utilize evidence-based conversational techniques such as Motivational Interviewing (MI)~\cite{mi-1} and Cognitive Behavioral Therapy (CBT)~\cite{cbt-1} to provide round-the-clock mental health support~\cite{Almusharraf,Sun2023VirtualSupport,linwei-mi,MI-chatbot,smoking-chatbot,therapy_cbt_llm,Survey-on-psychotherapy-chatbots}.
Traditionally, these agents have relied on rule-based approaches~\cite{rule-based-system} grounded in expert-crafted scripts~\cite{conversation_design_1,conversation_design_2} to ensure therapeutic adherence and safety. 
\revise{While effective, these approaches often produce rigid, non-adaptive conversations and require substantial expert effort to not only author dialogue content but also design the structured dialogue flows, as illustrated in Fig~\ref{fig:dataset_example}.}
Recent advances in Natural Language Generation (NLG)~\cite{nlg} have enabled more dynamic chatbot interactions. For example, MI chatbots have incorporated rephrasing and template-based generation\cite{reflective_rephrase} to improve engagement~\cite{Almusharraf,smoking-chatbot,domin2023verve}, while models trained on CBT data~\cite{therapy_cbt_llm} demonstrated the potential to deliver structured and principle-based therapeutic dialogues. 
Hybrid approaches~\cite{hybrid_ca,Plug_play} have emerged to integrate LLMs with rule-based systems, with the aim of combining generative flexibility with expert-guided structure and safety. 
However, these systems still require large domain-specific datasets, which are scarce in psychotherapy due to privacy concerns and data sensitivity.



The rise of large language models (LLMs)~\cite{llm_1} opens new opportunities for personalized, empathetic, and engaging digital psychotherapy~\cite{survey_llm_psychotherapy} and mental heath treatments~\cite{generative_AI_chatbot_mental_health}. 
Psychotherapy requires a real-time adaptation to subtle motivational, emotional, and behavioral cues.
\revise{Unlike simpler NLP models, advanced LLMs like ChatGPT~\cite{chatgpt} are capable of generating responses that are naturally fluent, contextually appropriate, and emotionally attuned.  Psychotherapeutic strategies such as reflective listening, a core element in MI and CBT, demand nuanced interpretation and reflective response, which requires far more than surface-level text manipulation such as paraphrasing or summarization. Furthermore, managing therapeutic dialogues involves pacing, topic transitions, goal alignment, and not merely following scripted dialogue flows.}
\camera{Yet, applying LLMs in the psychotherapeutic context to develop LLM-powered or hybrid~\cite{hybrid_ca} psychotherapy chatbots still poses three technical challenges}:
\revise{\textbf{(1)}} LLMs lack domain-specific knowledge to initiate and sustain conversations on specific \textit{psychotherapeutic topics} (see Fig~\ref{fig:dialogue_example}), such as increasing intrinsic motivation, or cognitive behavioral practices~\cite{cbt-1} like ``mindfulness''~\cite{mindfulness}, 
and cognitive distortions such as  ``should statements''~\cite{Should_Statements}.
\revise{\textbf{(2)}} LLMs also struggle to generate structured and goal-directed \textit{therapeutic questions} (see Fig~\ref{fig:dataset_example}) that guide behavioral interventions.
Moreover, the reliance on fully expert-crafted dialogue scripts presents another challenge: 
\revise{\textbf{(3)}} Creating these expert-crafted dialogue scripts is time-consuming and labor-intensive, posing barriers to scalability and broader adoption in real-world therapeutic contexts.


\begin{figure*}
\centering
\includegraphics[width=0.980\textwidth]{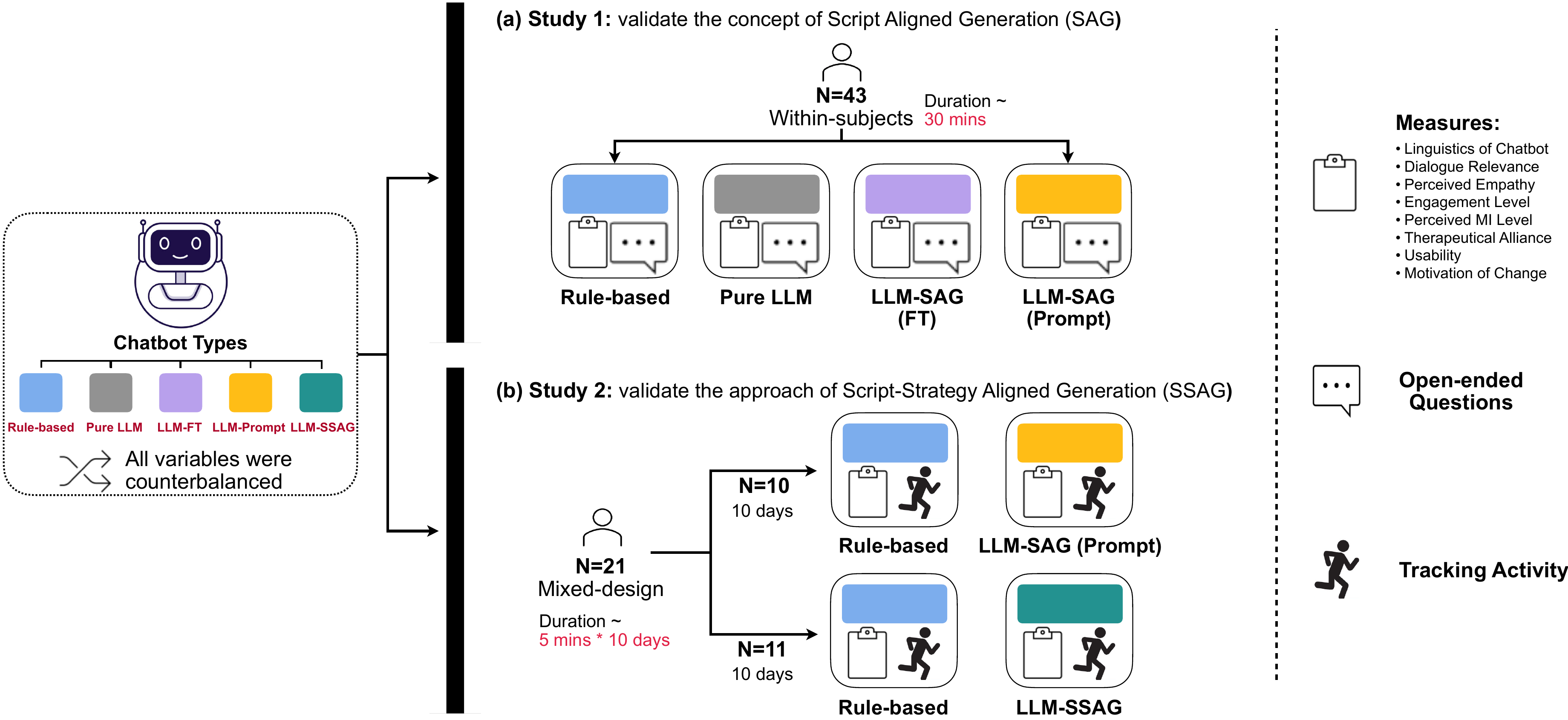}
\vspace{-1.6mm}
\caption{
The procedure and design of the two studies conducted to evaluate the chatbots in delivering the psychotherapy for health behavioral intervention
}
\vspace{-3.6mm}
\label{fig:procedure}
\end{figure*}

Despite the advancements of LLMs, \revise{it remains unclear whether expert-crafted dialogue scripts are still necessary to guide LLMs for psychotherapy, specifically in balancing generative flexibility with the structured and goal-oriented nature of evidence-based therapeutic conversations.}
\revise{This leads to our first research question. 
\textit{
RQ1: Do expert-crafted dialogue scripts remain essential for chatbot-delivered psychotherapy in the era of LLMs?}}
\revise{To explore this question, we propose Script-Aligned Generation (SAG), a concept that aligns LLMs with expert-crafted dialogue scripts to retain therapeutic structure while enabling conversational flexibility.}
To rigorously examine RQ1 and validate SAG, we define \revise{two sub-questions}:
\revise{\textit{RQ1.1: How do LLM-powered chatbots aligned with expert-crafted dialogue scripts compare to rule-based chatbots and pure LLMs for psychotherapy?}} and
\revise{\textit{RQ1.2: How do LLMs aligned via fine-tuning and prompting differ in performing psychotherapy?}}
To answer these questions and address the first two challenges, we conducted \revise{Study 1 (Fig~\ref{fig:procedure}.a)}, comparing four chatbot types:
a rule-based chatbot using static expert-crafted scripts;
a pure LLM without alignment;
and two LLM-powered chatbots aligned with expert-crafted scripts (i.e., SAG) via either fine-tuning or prompting.
Results showed that LLM-powered chatbots by SAG significantly outperformed both rule-based and pure LLMs across key assessing metrics, including linguistic quality, therapeutic relevance, empathy, engagement, MI adherence, and motivation enhancement. 
\revise{These findings highlighted the continuing importance of expert-crafted scripts in enabling LLMs to deliver safe and therapeutically effective psychotherapy.}


\revise{Although SAG aims to enable LLMs to balance conversational flexibility with therapeutic effectiveness by aligning them with expert-crafted dialogue scripts, it still relies on fully expert-crafted scripts, posing scalability and cost challenges for real-world psychotherapy applications.}
\revise{To address challenge (3), we proposed Script-Strategy Aligned Generation (SSAG), a more flexible and collaborative alignment approach. Unlike SAG, SSAG requires only partial expert input:} core psychotherapeutic topics, key therapeutic questions, and optional advice (see Fig~\ref{fig:ssag}). 
\revise{Drawing on prior work in CSCW~\cite{Therapeutic_Skills_Affect_Clinical,MI_Strategies_cscw} and NLP~\cite{cos_xsun} that employ therapeutic strategies to guide human therapists and LLMs for facilitating effective therapy,} SSAG enabled LLMs to dynamically generate dialogues aligned with these therapeutic strategies, such as asking questions, reflective listening, and giving advice, enabling alignments with both partial expert-crafted scripts and evidence-based therapeutic strategies~\cite{MI_Strategies_cscw, misc-1}.
We therefore ask our second research question:
\revise{\textit{RQ2: Can psychotherapy chatbots using SSAG achieve comparable conversational quality and therapeutic effectiveness to those using SAG?}}
To evaluate this, we conducted Study 2 (Fig~\ref{fig:procedure}.b), a 10-day field study comparing three chatbots: (1) a rule-based chatbot as baseline, (2) an SAG-aligned chatbot (via prompting) as in Study 1, and (3) an SSAG-aligned chatbot. 
\revise{Results showed that SSAG matched SAG in therapeutic effectiveness while sustaining empathy and engagement, suggesting that flexible alignment can deliver comparable outcomes with less expert scripts.}

\revise{
To distinguish between SAG and SSAG,} we defined SAG as strict alignment that relies on fully scripted dialogue content and flows, whereas SSAG is a flexible alignment approach that requires only partial expert input. SSAG enabled domain experts to contribute at a higher level of abstraction by specifying key therapeutic elements (e.g., psychotherapeutic topics and questions as illustrated in Fig~\ref{fig:ssag}) rather than scripting the entire dialogues.
This leads to our third research question:
\minor{\textit{RQ3: To what extent can SSAG reduce reliance on expert-scripted content for developing psychotherapy chatbots?}}
Findings from Study 2 showed that \revise{SSAG chatbots achieve comparable performance to those using SAG while reducing the need for fully expert-crafted dialogue scripts, positioning LLMs as co-authors of therapeutic dialogues.}
\revise{SSAG is situated within CSCW by facilitating collaboration among psychotherapy experts, developers, and LLMs. This helps enable the scalable, co-designed development of chatbots that have the potential to be both engaging and therapeutically effective for health behavioral interventions.}


\revise{We chose to focus on MI and CBT due to their strong evidence base and complementary structures.} MI, a flexible, client-centered approach, emphasizes reflective listening and open-ended questioning to enhance intrinsic motivation~\cite{mi-1,mi-2,miller428motivational,ALPERSTEIN2016393}, making it well-suited for evaluating the ability of LLMs to support engaging and context-aware conversations. CBT, \camera{by comparison}, offers structured, goal-oriented dialogues like cognitive reframing~\cite{cbt-1}, aligning well with LLMs guided by specific therapeutic objectives.
\revise{By combining MI’s flexibility with CBT’s structure, we can better explore whether aligned LLMs can uphold therapeutic adherence without sacrificing perceived conversational quality such as empathy, a tradeoff investigated in prior work about CBT~\cite{therapy_cbt_llm, Machine_and_Human_Understanding_of_Empathy}.}
\minor{Importantly, SSAG can be generalized beyond MI and CBT. By grounding generation in high-level therapeutic strategies, such as asking, reflecting, and advising, identified in prior CSCW work~\cite{MI_Strategies_cscw, Therapeutic_Skills_Affect_Clinical}, SSAG can be extended to other frameworks like Acceptance and Commitment Therapy (ACT)~\cite{act} or digital mental health coaching~\cite{dt_1, dt_2} by substituting context-appropriate, expert-defined strategies.}
This positions SSAG as a scalable, adaptable approach for a wide range of evidence-based digital health interventions.


\revise{This work contributes to CSCW by advancing the effective development of LLM-powered psychotherapy chatbots for health behavioral interventions.}
\revise{First,} we proposed and evaluated Script-Aligned Generation (SAG), a concept that aligns LLMs with fully expert-crafted scripts to balance therapeutic effectiveness and conversational flexibility, highlighting the ongoing importance of human expertise in guiding LLMs. 
\revise{Second,} we introduced Script-Strategy Aligned Generation (SSAG), a more flexible approach that \minor{reduces reliance on expert-scripted content by requiring only partial human input (e.g., therapeutic topics and questions), enabling efficient development of psychotherapy chatbots supported by LLMs.}
\revise{Third,} we contributed the first dataset of expert-authored psychotherapy dialogues grounded in evidence-based techniques (MI and CBT), supporting future research on LLM alignment for psychotherapy. 
\revise{Together, these contributions demonstrate that expert guidance remains critical for safe and effective digital psychotherapy, and that SSAG offers a practical path toward scalable, controllable, and efficient development of therapeutically aligned chatbots in the era of LLMs.}

%% file: 3-Related.tex
\section{Related Work}


\subsection{Chatbots and Conversational Design for Digital Psychotherapy and Behavioral Intervention}

The application of chatbots or conversational agents (CAs) in delivering psychotherapy and health- interventions has gained significant attention as a means of broadening access to psychotherapeutic and mental health support services like health behavioral intervention~\cite{Almusharraf,Sun2023VirtualSupport,smoking-chatbot} and mental healthcare~\cite{Survey-on-psychotherapy-chatbots,MI-chatbot}. 
Rule-based chatbots~\cite{rule-based-system}, which operate on predefined dialogue scripts, are widely used in digital psychotherapy for their high controllability and explainability, essential qualities in sensitive fields such as mental healthcare and psychotherapy. 
These systems ensure precision in addressing therapeutic needs and prevent deviations from clinically validated pathways.
For example, studies have used rule-based chatbots for Motivational Interviewing (MI)~\cite{mi-1,mi-2} and Cognitive Behavioral Therapy (CBT)~\cite{cbt-1} to support interventions like smoking cessation~\cite{Almusharraf,smoking-chatbot} and physical activity promotion~\cite{Sun2023VirtualSupport,article1}, where controlled, structured responses are crucial.

Despite the benefits of controllability and transparency, rule-based chatbots face limitations in dialogue empathy and flexibility. Their interactions tend to be rigid, lacking the dynamic adaptability of human therapists, which is particularly essential in psychotherapy, where a nuanced understanding of user queries and empathetic responses is critical~\cite{Therapeutic_Skills_Affect_Clinical}. 
For instance, a core principle of MI or CBT is reflective listening~\cite{misc-1}, where the therapist mirrors the client’s expressions to promote self-reflection and insight. However, rule-based chatbots struggle to emulate such responsive empathy~\cite{linwei-mi}, as their responses are confined to predefined scripts that may not fully capture the depth of user needs or emotional nuances.
More importantly, rule-based chatbots typically rely on intent-based dialogue systems~\cite{rule-based-system}, where domain experts design expert-crafted scripts to define chatbot's response to various user intents, ensuring consistent therapeutic guidance and simulating therapeutic conversations with a high degree of accuracy and reliability. 
However, this dependence on expert-written scripts makes rule-based chatbots highly resource-intensive, as they require extensive input from experts to develop domain-specific dialogue scripts~\cite{conversation_design_1,conversation_design_2,conversation_design_3}. 
Implementing these expert-crafted dialogues often involves encoding intricate conversation designs, including conversational content and flows, that incorporate strategies from psychotherapy such as MI or CBT, making the development process time-consuming and costly.
Therefore, our work created a dataset with extensive expert-crafted dialogue scripts tailored for physical activity interventions using MI and CBT.


\subsection{Generative Language Model for Digital Psychotherapy and Behavioral Intervention}

\revise{The integration of generative language models~\cite{survey_llm_psychotherapy,shen-etal-2020-counseling} into digital psychotherapy} shows promise in enhancing conversational flexibility and depth in chatbot-driven therapeutic interventions.
Initial efforts to improve rigid, rule-based chatbots led to hybrid systems from prior work by~\cite{hybrid_ca} that combined Natural Language Generation (NLG)~\cite{nlg} with predefined dialogue scripts. These systems aimed to improve conversational fluidity by incorporating generative elements to create more empathetic, reflective responses, a core principle of many psychotherapeutic techniques such as MI~\cite{mi-1,mi-2} and CBT~\cite{cbt-1}. For example, hybrid models generate reflective statements or paraphrases to encourage clients to explore their thoughts while staying aligned with therapeutic goals~\cite{Plug_play,reflective_rephrase,hybrid_ca}.
With advancements in large language models (LLMs)~\cite{llm_1,llm_2}, the potential for more adaptive and contextually aware dialogues in digital psychotherapy has grown considerably. LLMs offer a more powerful ability to generate varied and nuanced responses, which can help maintain engagement and improve the user experience during psychotherapy~\cite{cos_xsun,mi_reflection}. 
In MI or CBT, where client engagement and reflective listening are critical~\cite{linwei-mi}, LLMs have shown promise by producing responses that align with therapeutic techniques~\cite{therapy_cbt_llm,cos_xsun}, with a more flexible and engaging conversational style~\cite{PELAU2021106855,generative_AI_chatbot_mental_health,Therapeutic_Skills_Affect_Clinical}. 
A recent study~\cite{generative_AI_chatbot_mental_health} demonstrated the effectiveness of generative AI chatbots in providing clinical-level mental health treatments, supported by evidence from a long-term randomized controlled trial.

Despite these advantages, the use of LLMs in psychotherapy presents considerable challenges~\cite{llm_in_psychology}, especially concerning \revise{technical controllability, model transparency~\cite{30}, cultural understanding~\cite{therapy_cbt_llm} and ethical concerns~\cite{Ethical} in sensitive contexts}. 
Unlike rule-based systems that follow experts' pre-defined guidance or expertise, LLMs operate with a black-box nature that generates responses based on probabilistic models~\cite{probabilistic_model} that may lack explicit adherence to therapeutic principles \revise{or ethical standards~\cite{Ethical}}. 
This lack of predictability and controllability poses risks, as \revise{LLMs may inadvertently generate responses that are insensitive, ethically inappropriate, or even harmful, especially in sensitive interactions where mental health and emotional well-being are involved.} Furthermore, LLMs struggle to consistently apply psychotherapeutic counseling techniques, such as MI or CBT, as they lack the ability to systematically follow therapeutic strategies without expertise guidance. 
\revise{
These challenges have prompted further investigations into how LLMs and generative AI systems can be instructed or aligned with therapeutic principles~\cite{cos_xsun,welivita2023boosting,Therapeutic_Skills_Affect_Clinical}, human values~\cite{align_llm_human_reasoning,rlhf_language,rlhf}, external knowledge~\cite{deng2023knowledgeenhanced}, and ethical standards~\cite{align_llm_ethics_1,align_llm_ethics_2} to ensure safe and effective deployment in sensitive contexts such as psychotherapy for health intervention.
}


\subsection{\revise{Aligning LLMs with Domain Expertise and Human Instructions for Psychotherapy}}

Recent advances in aligning LLMs has evolved beyond linguistic fluency to emphasize adherence to goal-directed and domain-informed dialogue. 
This shift from purely probabilistic generative models~\cite{probabilistic_model} to more adaptive systems has introduced approaches such as instructed dialogue generation~\cite{InstructDial}, reinforcement learning from human feedback (RLHF)~\cite{rlhf_language, rlhf}, and context-aware generation~\cite{Context-Aware-Dialogue-Generation}.
These approaches enable LLMs to tailor response generation based on specific conversational strategies, user intent, and domain objectives~\cite{grounded_1, welivita2023boosting, cos_xsun}. 
This also enables more effective mixed-initiative dialogue~\cite{tu2022misc, deng2023knowledgeenhanced}, where both the user and model collaboratively guide the conversational interaction.

In psychotherapy, LLMs offer transformative potential to deliver flexible and engaging interventions; \revise{however, prior work by Iftikhar et al.~\cite{therapy_cbt_llm} indicates that instructed LLMs may overly rely on therapeutic techniques with compromising conversational qualities such as empathy.
To address this, alignment approaches should guide LLMs to balance conversational quality with adherence to structured therapeutic principles, which makes the alignment of LLMs with domain expertise critical to ensure safety, engaging, and goal-directed conversations.}
Aligned dialogue generation is expected to support models to embed psychological and empathetic principles within responses~\cite{generative_AI_chatbot_mental_health, Human_AI_collaboration_enables_more_empathic, cos, Singhal2023}, allowing LLMs to effectively emulate the precision of rule-based systems while retaining the flexibility needed for nuanced, context-sensitive support in psychotherapeutic settings.

\revise{Two dominant alignment approaches have emerged: fine-tuning and prompting. 
Fine-tuning~\cite{fine_tuning} integrates domain expertise directly into model weights, offering strong domain fidelity~\cite{welivita2023boosting, generative_AI_chatbot_mental_health, therapy_cbt_llm}. However, it is resource-intensive and dependent on large, sensitive datasets that are difficult to obtain in psychotherapy. Prompting~\cite{prompt_methods}, especially with in-context learning methods like Chain-of-Thought~\cite{cot_1, cot_2, cue-cot} and Tree-of-Thoughts~\cite{tree_of_thoughts}, provides a more scalable and flexible alternative, guiding LLM outputs with structured cues instead of continuous model retraining. 
Recent work~\cite{cos_xsun, basar-etal-2025-well} has shown that incorporating therapeutic strategies into prompts allows LLMs to produce more therapeutic-aligned responses~\cite{Therapeutic_Skills_Affect_Clinical,MI_Strategies_cscw}. 
Comparative studies~\cite{ft_prompt_rag_llm_mental_health} also validate both prompting and fine-tuning as effective approaches for mental health tasks, supporting their potential in psychotherapeutic interventions.}

Despite this progress, \revise{these approaches typically assume access to comprehensive domain-specific data for aligning the LLMs, which can be costly and time-consuming for domain experts to produce. 
To address this, we proposed Script-Strategy Aligned Generation (SSAG) in this work, an alignment approach that requires only partial expert-crafted dialogue scripts, thereby reducing reliance on fully scripted dialogue content or flows, while using LLMs to dynamically manage dialogue flow and generate responses. SSAG supports a collaborative expert-LLM co-authoring workflows and enables a controllable alignment through the stepwise therapeutic strategy prediction mechanism. Compared to prompting and fine-tuning, SSAG offers a more flexible and efficient alternative for the real-world development of psychotherapy chatbots, especially in the contexts where data is limited and expert labor is costly. 
}


%% file: 4-Dataset.tex
\section{Creating Dataset with Expert-Crafted Dialogue Scripts}
\label{dataset}

\begin{figure*}[!ht]
\centering
\includegraphics[width=0.992\textwidth]{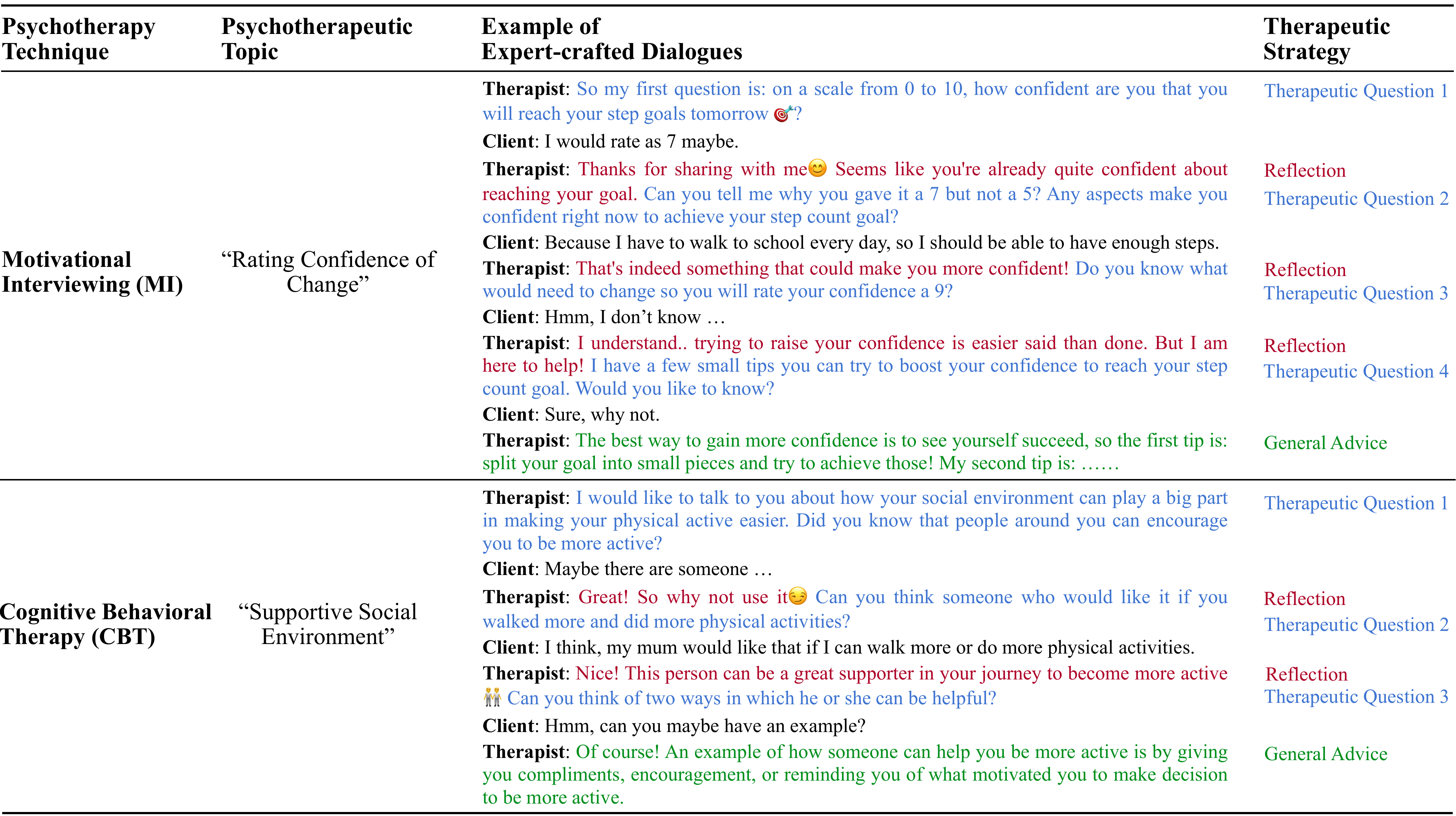}
\vspace{-2.0mm}
\caption{Examples of expert-crafted dialogues in motivational interviewing (MI) and cognitive behavioral therapy (CBT) with different psychotherapeutic topics. \revise{The dialogues contain 1) psychotherapeutic topics, 2) therapeutic questions, 3) reflections, 4) general advice, and 5) dialogue flows pre-designed as in Fig~\ref{fig:dataset_example}.}}
\label{fig:dialogue_example}
\vspace{-3.6mm}
\end{figure*}

\camera{A team of fifteen experts (i.e., research associates, each holding a BSc or higher degree in clinical or health psychology),} collaboratively developed a comprehensive dataset of pre-scripted, tree-structured dialogues designed to support physical activity interventions. 
\revise{Pre-scripted dialogues include both dialogue content and flows in tree-structure as shown in Fig~\ref{fig:dataset_example}.}
Grounded in the \revise{psychotherapeutic techniques of MI~\cite{mi-1,mi-2} and CBT~\cite{cbt-1},} these scripts target common therapeutic scenarios in behavioral interventions. 
Examples of crafted dialogues are shown in Fig~\ref{fig:dialogue_example}, covering topics such as ``rating confidence of change'' in MI and ``supportive social environment'' in CBT.

The creation process began with brainstorming sessions, where experts identified key psychotherapeutic topics relevant to behavioral interventions, focusing on physical activity. 
Topics like overcoming exercise barriers and improving sleep hygiene were chosen to cover diverse intervention scenarios. 
Each expert then crafted dialogues for specific topics, ensuring the dataset captured the depth and variability of real therapy while adhering to MI and CBT principles.
To ensure clarity and authenticity, a think-aloud protocol~\cite{think_aloud} was used. 


The final dataset includes over 1,800 expert-written utterances across 26 \textbf{psychotherapeutic topics} as detailed in~\ref{appendix_dialogue}.
\revise{
The dataset includes a mix of closed- and open-ended questions, reflections, and general advice, key elements for eliciting client self-reflection and promoting engagement, which are core objectives of both MI and CBT.}
Each topic is structured by a series of \textbf{therapeutic questions}.
The tree-structured format as shown in Fig~\ref{fig:dataset_example} ensures a coherent flow and branching, designed to mirror real-world psychotherapeutic dynamics. 
\revise{
In order to support reproducibility of our work, we aim to make the dataset available upon request, with the creation of a data sharing agreement following EU intellectual property management~\cite{eu_ip_management} guidelines and the respective institute wherein this data belongs.
More detailed dialogue examples are provided in~\ref{appendix_dialogue_mi} and \ref{appendix_dialogue_cbt}
}

%% file: 5-Study-1-Methods.tex
\section{Study 1: Concept of Aligning LLM with Full Expert-Crafted Dialogue Scripts}

\subsection{\revise{Concept Validation}: Script-Aligned Generation (SAG) by Fine-tuning and Prompting}

\revise{
Study 1 aimed to validate the concept of Script-Aligned Generation (SAG), which aligns LLMs with full expert-crafted dialogue scripts for psychotherapy. 
SAG seeks to balance therapeutic fidelity with the conversational flexibility and engagement inherent in LLMs. 
The goal of Study 1 was to assess SAG’s potential to enhance LLMs’ potential to deliver flexible, engaging, and expert-guided psychotherapy.
To implement the concept of SAG, we proposed two alignment approaches: fine-tuning (LLM-SAG (FT))~\cite{fine_tuning} and prompting (LLM-SAG (Prompt))~\cite{prompt_methods}, to align LLMs with expert-crafted, tree-structured dialogue scripts designed for health intervention as shown in Fig~\ref{fig:dataset_example},} aiming to determine which provides the best combination of therapeutic effectiveness and conversational qualities.

\begin{figure*}[!ht]
\centering
\includegraphics[width=0.92\textwidth]{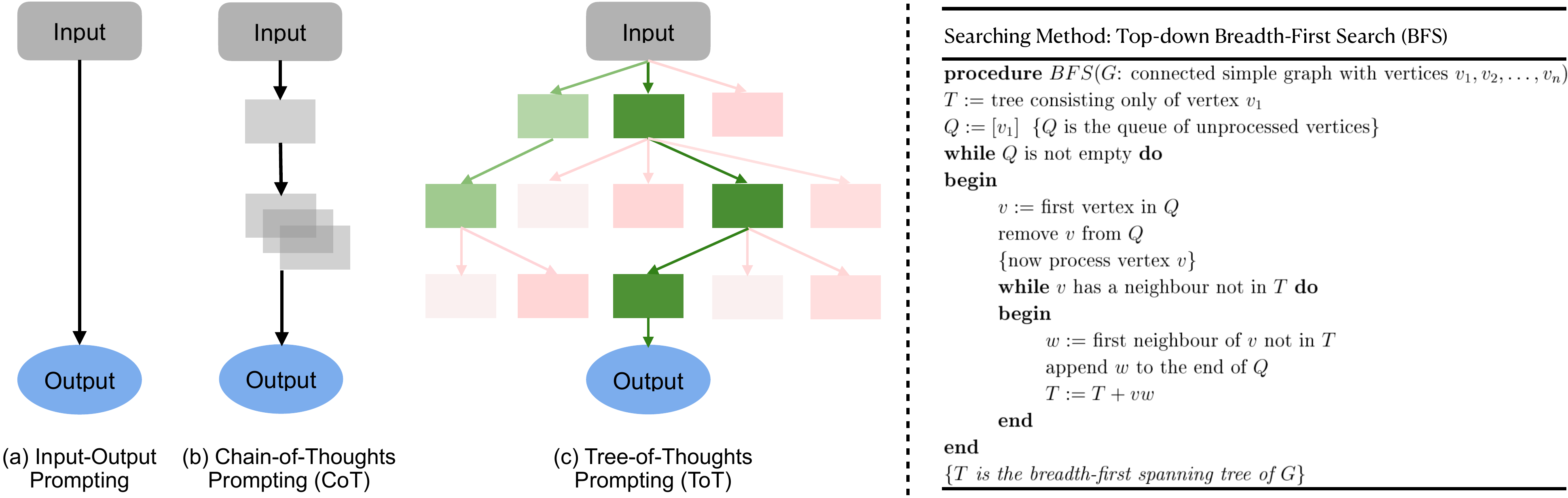}
\caption{Visualization of aligning LLMs with expert-crafted tree-structured dialogue scripts via Tree-of-Thoughts prompting~\cite{tree_of_thoughts}.}
\label{fig:tot}
\vspace{-3.6mm}
\end{figure*}


\subsection{Study Methods}

\subsubsection{Study design and procedure}
\camera{A comprehensive overview of the study procedure is illustrated in Fig~\ref{fig:procedure} (a).}
We conducted an evaluation study employing a within-subjects design to compare four chatbot types: 
1) a rule-based chatbot (``rule-based'') strictly followed the expert-crafted dialogues;
2) a pure LLM (``pure LLM'') \revise{without any alignment with expert-crafted dialogues;}
and LLM-powered chatbots strictly aligned with the \revise{full} expert-crafted scripts through either 3) prompting (``LLM-SAG (Prompt)'') or 4) fine-tuning (``LLM-SAG (FT)'').

Before the study, each participant received an information letter and provided informed consent.
After, participants interacted with each chatbot in \revise{counterbalanced} orders and completed a survey immediately after interaction to assess their experience with that specific chatbot. 
This within-subjects design allowed for a direct comparison of user assessments across chatbot types under consistent conditions, aiming to identify which type of chatbot best performs in facilitating the therapeutic interactions.


\subsubsection{Types of chatbots}
\label{study_1_chatbots}
Study 1 examined the following four types of chatbot: 

\noindent 1) \textbf{rule-based chatbot.}
\revise{
This chatbot was implemented using the RASA framework~\cite{rasa}, a popular open-source platform for building rule-based chatbots. It strictly followed the expert-crafted dialogue scripts detailed in Section~\ref{dataset}, using predefined intents, responses, and dialogue flow rules encoded in RASA’s dialogue management model~\cite{rasa}. 
This chatbot did not generate responses beyond what is explicitly pre-scripted, ensuring high controllability. 
This condition served as a baseline, representing a traditional chatbot approach where all conversational paths are predetermined by experts.
}

\noindent 2) \textbf{pure LLM.}
\revise{
This chatbot used a GPT-4o model~\cite{gpt4} without alignment to expert-crafted dialogue scripts. At the start of each session, the LLM was prompted with:
``You are a psychotherapist conducting a session to promote healthier behavior using Motivational Interviewing (or Cognitive Behavioral Therapy). The current therapeutic topic is [given topic]. The definition of this topic is [description of the given topic].''
The topic was selected from a set of expert-defined \textbf{psychotherapeutic topics} (as demonstrated in Section~\ref{dataset} and detailed in~\ref{appendix_dialogue}). Participants can interact with this chatbot freely within the specified topic. This setup evaluated the pure LLM's performance in delivering psychotherapy with high-level instructions, but without script guidance or alignment.}


\noindent 3) \textbf{LLM-SAG (FT).}
\revise{
We fine-tuned a GPT-4o model using our expert-crafted dialogue dataset, which consisted of tree-structured dialogues authored and reviewed by health psychology experts for clarity and coherence (see Section~\ref{dataset}). 
We employed supervised fine-tuning involving input-output pairs, ensuring the LLM’s responses aligned with expert-crafted dialogue responses and flows.}
Additionally, to evaluate the \revise{chatbot’s adherence to the predefined dialogue flow, participants can ask “out-of-script” questions to test the chatbot's ability to respond appropriately while maintaining alignment with the specific dialogue flow. 
This method was also applied to the LLM-SAG (Prompt) chatbot.}

\noindent 4) \textbf{LLM-SAG (Prompt).}
\revise{For a more scalable and computationally efficient alternative to fine-tuning~\cite{when_to_ft}, we implemented a prompting-based alignment using the Tree-of-Thoughts (ToT) technique~\cite{tree_of_thoughts}, as visualized in Fig~\ref{fig:tot}. 
ToT guided the GPT-4o model through our pre-authored, tree-structured dialogue scripts using a Breadth-First Search (BFS) algorithm~\cite{bfs}. 
}
\revise{
We chose ToT for its effectiveness to align LLM with tree-structured expert-crafted dialogue scripts. 
\minor{Unlike other prompting techniques (e.g., few-shot or chain-of-thought~\cite{cot_1,cot_2}), ToT ensures consistency in multi-turn dialogue scenarios by enabling controlled, step-by-step navigation through predefined tree-based dialogue branches.}
At each turn, the LLM was prompted with the current dialogue context and a constrained set of valid next-turn response options through predefined dialogue branches in the tree structure, preserving the dialogue logic and therapeutic progression embedded in expert scripts.
Compared to fine-tuning, ToT is a more lightweight and flexible approach that avoids retraining while maintaining integrity to expert scripts.}
\minor{While ToT was well-suited for aligning with tree-structured scripts in this work, we did not compare it with alternative prompting strategies (e.g., chain-of-thought). Future work could evaluate how different prompting strategies perform in this context.}

\revise{
The prompting templates used for the LLM-powered chatbots are provided in Appendix B. 
The implementation details of the models and settings are presented in~\ref{appendix:implement_details}.}

\subsubsection{Web-based interfaces for chatbot interaction and evaluation}
\begin{figure*}[htbp]
\centering
\includegraphics[width=0.986\textwidth]{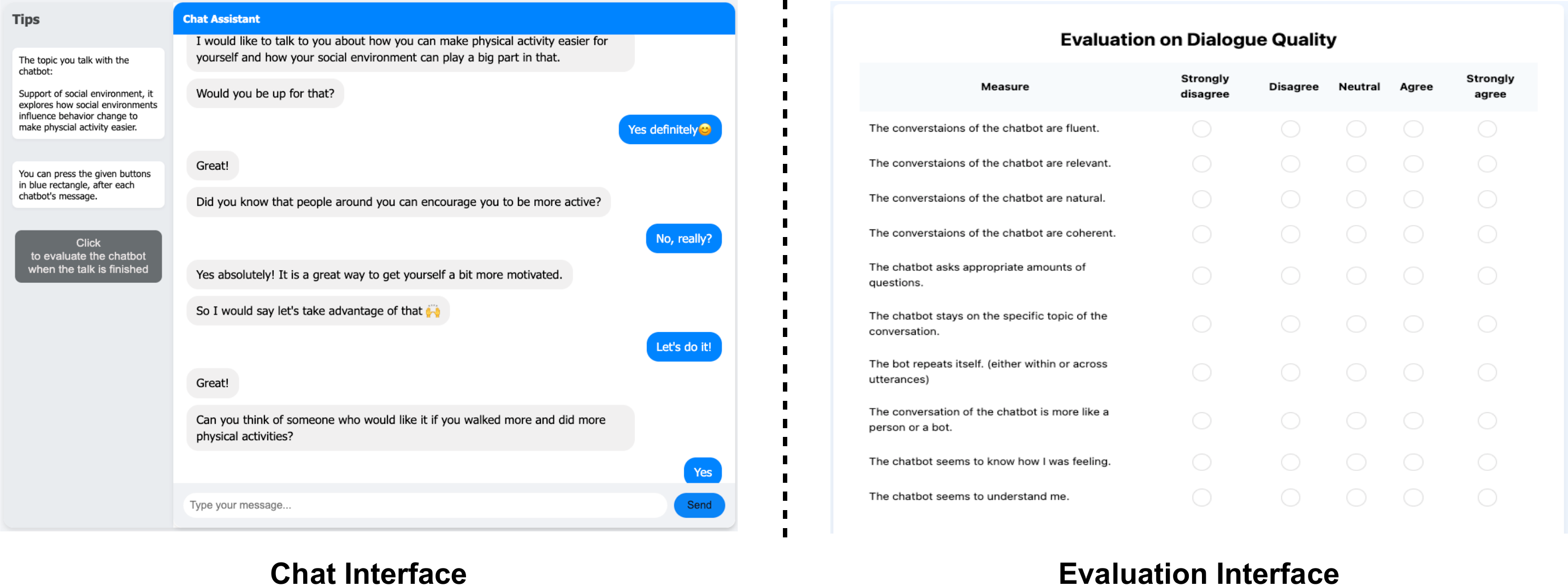}
\vspace{-2mm}
\caption{Web-based interfaces were developed for Study 1 to enable chatbot interactions and facilitate immediate evaluation.}
\label{fig:interface}
\vspace{-3.0mm}
\end{figure*}

The study was conducted using a self-developed web-based application that integrated both chatbot interaction and an evaluation survey into one seamless interface, as shown in Fig~\ref{fig:interface}. 
This design allowed participants to interact with the chatbot and provided immediate feedback in a continuous session, streamlining user experience. 
The interface was optimized for simplicity, minimizing distractions and ensuring smooth navigation, enabling participants to focus on their interactions without difficulties.

\subsubsection{Measures}
We assessed above four chatbot types using the measures as follows. 
All quantitative self-report measures used a 5-point Likert scale ranging from 1 (Strongly Disagree) to 5 (Strongly Agree). 

\noindent \textbf{Linguistic Quality}: Evaluated the fluency, relevance, naturalness, coherence, and the human-likeness of the chatbot's responses, using five assessing items adopted from ~\cite{nlg_measure_1,nlg_measure_2}. Example items are ``The conversations of the chatbot are fluent.'' and ``The conversation of the chatbot is more like a person rather than a bot.''

\noindent \textbf{Dialogue Relevance}: Inspired by work~\cite{nlg_measure_relevance}, two self-constructed items were used to specifically assess the alignment of expert-crafted therapeutic dialogues. Example items are ``The chatbot asks an appropriate amount of questions.'' and ``The chatbot stays on the specific topic of the conversation.''

\noindent \textbf{Empathy and Engagement}: Gauged perceived empathy of the chatbot's conversations, which is crucial for therapeutic effectiveness. We adopted the questionnaire from ~\cite{empathy_measure,linwei-mi} with two items: ``The chatbot seems to know how I felt.'' and ``The chatbot seems to understand me.''
To assess the chatbot's ability to maintain user interest and interaction as the engagement level, a questionnaire from ~\cite{engagement_measure} was adopted. Example items are ``I lost myself in the interaction with this chatbot.'' and ``The chatbot is enjoyable to talk to.''

\noindent \textbf{Perceived Motivational Interviewing (MI) Adherence}: Evaluated how well the chatbot simulates an MI session and adheres to the MI principles. We adopted a questionnaire from ~\cite{mi_measure,linwei-mi} with four items. Example items are ``The chatbot helped me talk about changing my behavior.'' and ``The chatbot helped me feel hopeful about changing my behavior.''

\noindent \textbf{Motivation for Change}: Measured the chatbot's impact on participants' motivation to change, a key therapeutic outcome. We self-constructed an item ``I am motivated to make changes in the behavior after I talk with the chatbot.''

\noindent \textbf{Therapeutic Alliance}: Explored the rapport and connection between the chatbot and the user. We adopted the questionnaire from~\cite{dwai} with six items. Example items are ``I believe this chatbot can help me to address my problem.'' and ``This chatbot encourages me to accomplish tasks and make progress for the change we discussed.''

\noindent \textbf{Usability}: Measured the ease of interaction with the chatbots. We adopted the questionnaire called Bot Usability Scale from~\cite{bus} with seven items. Example items are ``Communicating with the chatbot was clear.'' and ``The chatbot's responses were easy to understand.''

\noindent \textbf{Two Open-Ended Questions} were included to capture participants' feelings and opinions about chatbots: ``What have you enjoyed most or least about interacting with this chatbot?'' and ``What could be improved about this chatbot?''

\noindent \textbf{Automatic Evaluation Metrics}:
\label{auto_metrics}
We also developed two self-defined, tailored metrics to evaluate the chatbots' performance in delivering expert-crafted \textit{psychotherapeutic topics} and \textit{therapeutic questions} (shown in Fig~\ref{fig:dataset_example} and Fig~\ref{fig:dialogue_example}).
\textit{``Auto-Metric 1''} assesses the number of psychotherapeutic topics that reached a natural conclusion, reflecting the level of user engagement with chatbots. 
\textit{``Auto-Metric 2''} measures the total number of expert-crafted therapeutic questions asked by the chatbot, indicating its effectiveness in guiding the conversation toward specified therapeutic goals.


\subsubsection{Participants}
A power analysis using G*Power~\cite{gpower} indicated that at least 30 participants were required to detect a medium effect size ($d$ = 0.25) and \minor{$\alpha = 0.05$} with 90\% power. 
To ensure robustness, we recruited 43 participants (N=43) through institutional channels and social media. 
Participants had a diverse demographic profile, including variations in age, gender, and educational background. 
Eligibility criteria required participants to be at least 18 years old and fluent in English. 
Participation was voluntary, and participants were compensated with money or study credits for completing a 30-minute online session. 
This study was approved by the institutional ethics committee. 
Participant demographics are shown in Table~\ref{table:demographics}.

\vspace{-1.6mm}

\input{tables/demographics}


\subsubsection{Data analysis}
To analyze the impact of different chatbots on participants’ perceptions of psychotherapeutic interventions, we first tested the data's suitability for statistical analysis. Normality was assessed using the Shapiro-Wilk test~\cite{SHAPIRO1965}, and homogeneity of variance via Bartlett’s test~\cite{Arsham2011}. 
As the data violated normality assumptions, we applied a Generalized Estimating Equation (GEE) model~\cite{gee} to compare mean scores across chatbot conditions. 
Post-hoc pairwise comparisons between the LLM-SAG (Prompt) and LLM-SAG (FT) chatbots were conducted using Wilcoxon signed-rank tests~\cite{wilcoxon} with Bonferroni correction~\cite{bonferroni}. 

For the qualitative data, we conducted an inductive content analysis~\cite{elo2008qualitative} of responses to two open-ended questions. 
The first two authors developed an initial codebook using ATLAS.ti~\cite{atlas_ti} based on participants' perceptions and suggested improvements for each chatbot type. 
Both coders independently coded the responses, revising the codebook as new themes emerged. 
Codes were refined, merged where appropriate, and re-coded to ensure consistency. 

%% file: tables/demographics.tex
\begin{table}[!h]
\centering
\footnotesize
\renewcommand{\arraystretch}{0.98}
\begin{tabularx}{\columnwidth}{p{4cm} >{\raggedright\arraybackslash}p{7cm} >{\raggedright\arraybackslash}X}
\toprule
\textbf{Demographic} & \textbf{Categories} & \textbf{Numbers of Participants (\%)} \\
\midrule
Gender & Female & 22 (51.2\%)
\\ 
 & Male & 21 (48.8\%)
\\
\hline
Age & 18-24 & 23 (53.5\%)
\\
 & 25-34 & 14 (32.6\%)
\\
 & 35-44 & 2 (4.7\%)
\\
 & 45-54 & 3 (7.0\%)
\\
 & 65+ & 1 (2.3\%)
\\
\hline
Education & High school degree or equivalent & 8 (18.6\%)
\\
 & Bachelor’s degree & 15 (34.9\%)
\\
 & Master’s degree & 16 (37.2\%)
\\
 & Doctorate or higher & 4 (9.3\%)
\\
\hline
Professional Domain & Health and Medical Science & 7 (16.3\%)
\\
 & Science, Technology, Engineering, Mathematics (STEM) & 14 (32.6\%)
\\
 & Business, Economics, and Law & 6 (14.0\%)
\\
 & Communication, Arts, Culture and Entertainment & 5 (11.6\%)
\\
 & Education and Social Science & 8 (18.6\%)
\\
 & Government and Public Sector & 2 (4.7\%)
\\
& Other & 1 (2.3\%)
\\
\bottomrule
\end{tabularx}
\caption{Characteristics of participants in Study 1.}
\vspace{-7.6mm}
\label{table:demographics}
\end{table}

%% file: 5-Study-1-Results.tex

\subsection{Quantitative Findings}

\subsubsection{Automatic evaluation metrics}

Table~\ref{tab:res_automatic} shows the results of the automatic evaluation metrics, as demonstrated in Section~\ref{auto_metrics}. 
For ``Auto-Metric 1'' (i.e., measuring completion of each psychotherapeutic topic), the rule-based chatbot scores the highest, followed by LLM-SAG (Prompt) and pure LLM, which have similar scores. In ``Auto-Metric 2'' (i.e., measuring therapeutic questions asked), the rule-based chatbot again leads, with LLM-SAG (Prompt) close behind. The pure LLM scores the lowest, indicating a significant deviation from the expert-crafted dialogue scripts.

\input{tables/analysis_results_automatic}


\subsubsection{Descriptive statistics}

\revise{Participants reported moderate prior exposure to chatbots.}
\camera{The average frequency of chatbot use was 3.14 (SD=0.83), prior experience was 3.23 (SD=1.08), familiarity with chatbots was 3.44 (SD=0.73), and overall attitude toward chatbots was 3.40 (SD=0.79).}


Across the four chatbot types: rule-based, pure LLM, LLM-SAG (FT), and LLM-SAG (Prompt), the LLM-SAG (Prompt) consistently outperformed others across multiple measures. 
As shown in Table~\ref{tab:study_1_gee_rule_vs_others}, LLM-SAG (Prompt) achieved the highest ratings in linguistic quality (M=3.95, SD=0.23), dialogue relevance (M=3.39, SD=0.68), and empathy (M=3.07, SD=0.16), and showed strong performance in engagement (M=2.90, SD=0.33), perceived MI adherence (M=3.74, SD=0.13), motivation (M=3.42, SD=0.00), usability (M=3.82, SD=0.26), and therapeutic alliance (M=3.58, SD=0.46), surpassing other chatbot types. 
This indicated that it not only followed expert-crafted dialogue scripts closely but also delivered a more flexible and effective therapeutic experience overall.

\vspace{-1.6mm}

\input{tables/study_1_gee_rule_vs_others}
\vspace{-1.6mm}


\subsubsection{Rule-based vs. LLM-powered chatbots}

\begin{figure*}[!h]
\centering
\includegraphics[width=0.986\textwidth]{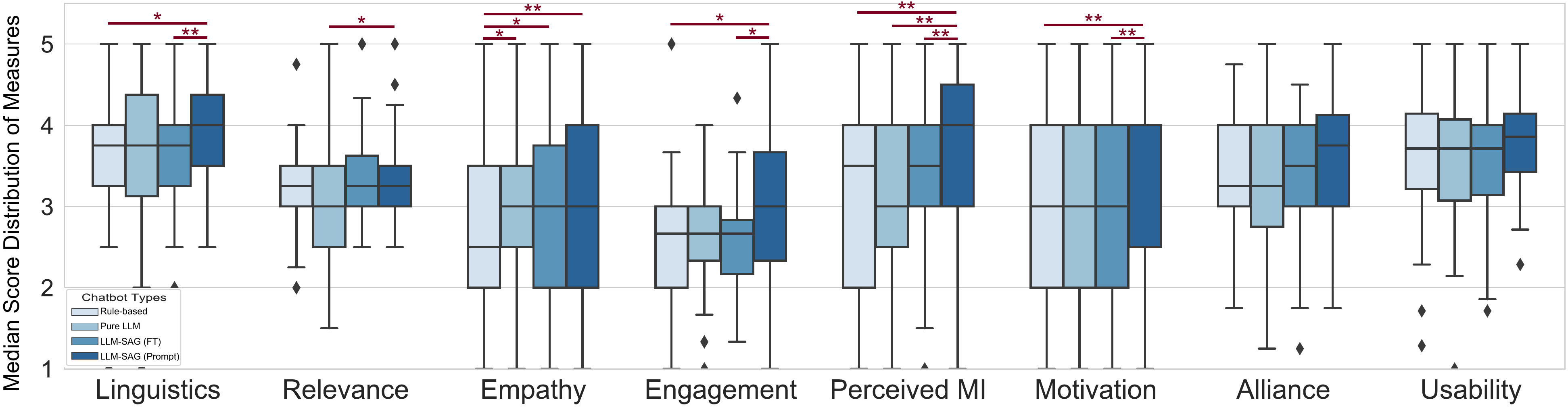}
\vspace{-1mm}
\caption{Distribution of median value of measures in Study 1 across four chatbot types.
\camera{Y-axis ranges from 1 to 5 (Likert scale). Each box shows the interquartile range, with the midline indicating the median. Dots outside represent outliers. Horizontal red lines with asterisks indicate statistically significant differences between conditions (*p < .05, **p < .01).}
}
\vspace{-2mm}
\label{fig:gee_analysis}
\end{figure*}

\revise{To address RQ1.1}, we compared the performance of rule-based and LLM-powered chatbots, including pure LLM, LLM-SAG (FT) and LLM-SAG (Prompt). 
As depicted in Table~\ref{tab:study_1_gee_rule_vs_others} and Fig~\ref{fig:gee_analysis}, LLM-SAG (Prompt) significantly outperformed the rule-based chatbot in linguistic quality (p=0.02), empathy (p=0.01), engagement (p=0.03), perceived MI adherence (p=0.01), and motivation change (p=0.01).
Pure LLM and LLM-SAG (FT) chatbots only showed significantly higher empathy ratings than the rule-based chatbot (p=0.04), but performed similarly or even worse in dialogue relevance and MI adherence.
LLM-SAG (Prompt) chatbot scored significantly higher than pure LLM in dialogue relevance and perceived MI, shown in Fig~\ref{fig:gee_analysis}. 
Moreover, all chatbots performed similarly in usability and therapeutic alliance.
These findings affirmed the value of aligning LLMs with expert-crafted scripts to preserve therapeutic quality and conversational engagement.


\subsubsection{Prompting (LLM-SAG (Prompt)) vs. Fine-tuning (LLM-SAG (FT))}

\begin{figure*}[!htbp]
\vspace{-0mm}
\centering
\includegraphics[width=0.53\textwidth]{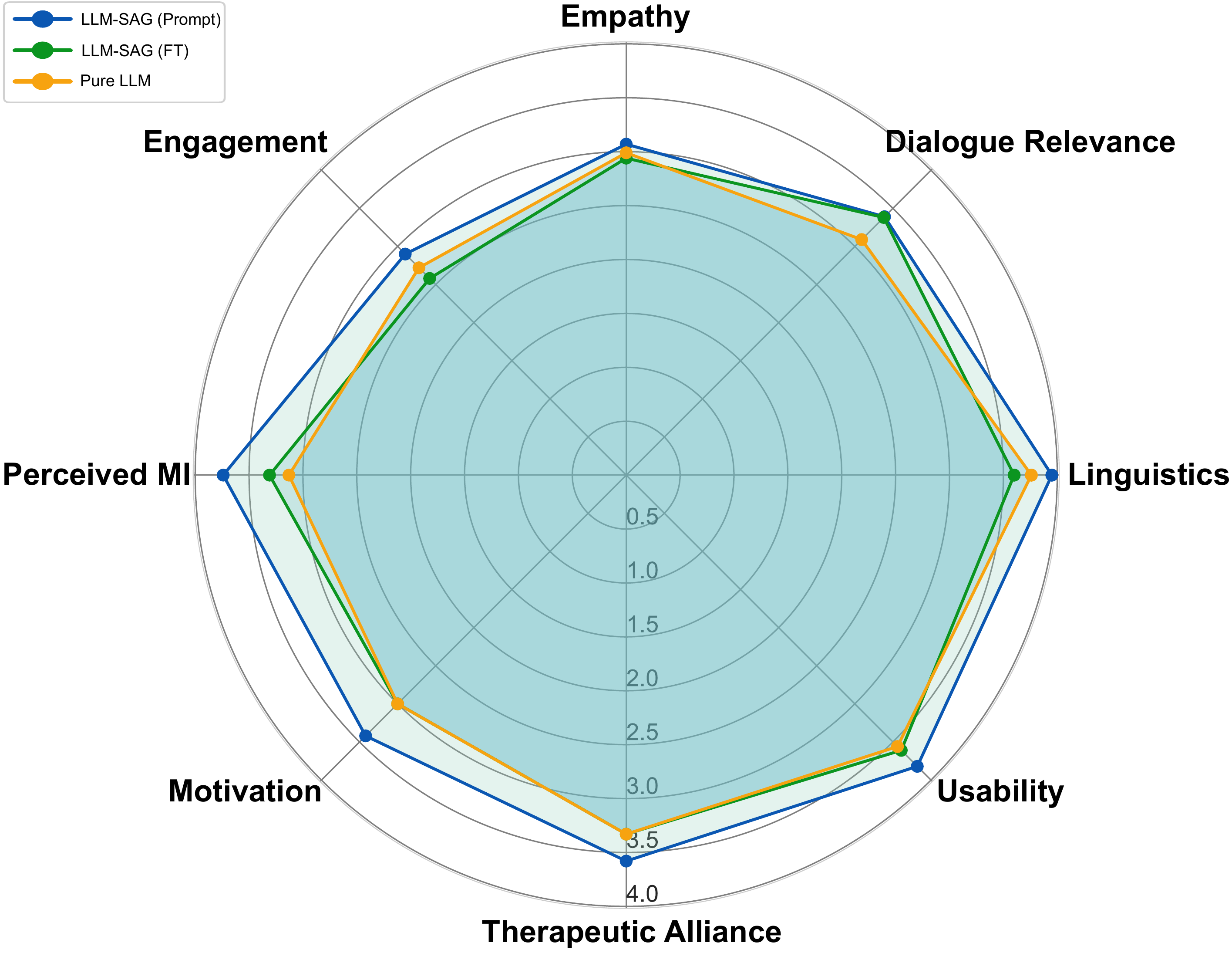}
\caption{Visualization of the comparison (mean value) between the pure LLM chatbot and LLM-SAG chatbots through prompting (LLM-SAG (Prompt) and fine-tuning (LLM-SAG (FT).
\camera{Each axis represents one of eight evaluation dimensions, with values plotted on a radial scale from 0 to 5 reflecting mean scores for each chatbot type.}
}
\vspace{-2.0mm}
\label{study_1_radar}
\end{figure*}

\revise{To address RQ1.2}, we compared two alignment approaches used to implement the concept of SAG: fine-tuning (LLM-SAG (FT)) and prompting (LLM-SAG (Prompt)).
Results in Table~\ref{tab:analysis_results_2} and Fig~\ref{study_1_radar} showed that LLM-SAG (Prompt) significantly outperformed LLM-SAG (FT) in assessing diemensions, including linguistic quality (p=0.01), engagement (p=0.02), perceived MI (p=0.01), and motivation change (p=0.02).
\revise{These findings suggested that prompting provides a more effective and flexible alignment approach for psychotherapy, likely due to its ability to flexibly navigate tree-structured dialogues while preserving expert-scripted dialogues for effective therapeutic experience, whereas fine-tuning may struggle to balance the adherence to the pre-crafted complex dialogue scripts and the adaptability to the ``out-of-scripts'' inputs from the participants.}

\vspace{-0.0mm}

\input{tables/study_1_wilcon}
\vspace{-3.6mm}

\subsection{Qualitative Findings: Open-ended Questions}


A total of 344 free-text responses were collected in Study 1 from 43 participants \revise{across four chatbot types (i.e, rule-based vs. pure LLM vs. LLM-SAG (FT) vs LLM-SAG (Prompt)).} 
\camera{298 non-empty responses were included for analysis.}
Findings revealed four key themes shaping participant perceptions of chatbots in psychotherapy for health behavioral intervention that \revise{specifically addressing RQ1}: 
\revise{(1) conversational quality and effectiveness, (2) balancing engagement with therapeutic adherence, (3) LLM-powered intelligences for digital psychotherapy, and (4) suggestions for improving chatbot-delivered psychotherapy.}


\subsubsection{\textbf{\revise{Effective chatbot-delivered psychotherapy through clarity, relevance and coherence}}}
\label{qualititive:effective_communication}

In therapeutic contexts, the effectiveness of chatbot interactions depends heavily on two intertwined factors: clarity and coherence in conversations, and the therapeutic relevance of the content. This theme highlights the need for chatbots to respond in a way that is easy to understand, logically structured, and directly aligned with users’ therapeutic goals.

The qualitative analysis showed that participants \camera{responded favorably to the rule-based chatbots} for their straightforward responses that were built on expert-crafted dialogue scripts. These chatbots were seen as effective in guiding users through therapeutic content with clarity. One participant remarked, “I liked the fact that the chatbot was really straightforward and clear in providing information, leading the users into questions. (rule-based chatbot)”
However, participants also highlighted instances where clarity was lacking, especially in pure LLM-based chatbots. One participant shared, “It is not clear, and sometimes I don't know what I should say. Some examples are not clear at all. I wanted to see real instances, but it said ambiguously. (pure LLM)” This feedback highlights the need for clearer instructions, concrete examples, and more direct conversational guidance.

Therapeutic and contextual relevance were also viewed as essential. Some participants felt pure LLMs lacked focus, as one noted: “The topic is not quite relevant to the topic given; it would be good if it gave more specific advice on the topic we talked about for helping me in behavioral change.” This highlights the importance of responses grounded in therapeutic objectives and user needs.
While rule-based chatbots were appreciated for clarity, their rigidity was a frequent criticism. One participant noted, “I do not like chatbots with pre-set responses. There was no flexibility in what I could talk about.” This inflexibility was seen as a barrier to deeper engagement. In contrast, LLM-based chatbots, particularly those aligned with expert-guided prompts (LLM-SAG), were praised for their adaptability and contextual responsiveness. As one participant described, “The chatbot can understand what I talk about and always stay within the topic given. It also asks relevant questions to move the talk forward. (LLM-SAG (Prompt))”

Maintaining logical flow and staying on topic was viewed as essential of effective interventions. 
Participants valued when chatbots responded coherently. One participant noted the improved coherence in LLM-SAG chatbots compared to rule-based one: “I can feel the dialogue from the (rule-based) chatbot is similar to the previous one, and this time it (LLM-SAG (Prompt)) can understand what I said and give correct reply. I can feel it quite encouraging for change.”
Another participant emphasized the capability of aligned LLMs to stay on topic: “It guides my behavior change step by step in human-like conversations. And it is not overwhelming. It can understand my specific questions and give correct answers, and then it can go back to previous talk after my questions, interesting. (LLM-SAG (Prompt))”




\subsubsection{\textbf{\revise{Balancing conversational engagement and therapeutic adherence: Flexible vs. Structured}}}
\label{qualititive:tones}

Conversational style plays a pivotal role in shaping users’ perceptions and relationships with chatbots. Participants consistently highlighted how conversational style impacts their sense of engagement and therapeutic effectiveness. 

The rule-based chatbot provided well-structured dialogues for psychotherapeutic intervention following expert-crafted content and was noted as, "I love it guiding my behavior change step by step. Instead of directly asking me to do certain things, it guide me to think about how to deliver such changes."
However, participants described LLM-powered chatbots as more relatable and engaging than rule-based ones. The LLM-powered chatbots’ ability to mimic natural conversation made interactions feel natural and fluid. As one participant noted, “The tone of this chatbot is interesting and appealing. It talks like a human being. (LLM-SAG (FT))” In contrast, rule-based chatbots with their rigid, pre-set responses were disliked in this context as, "I do not like (rule-based) chatbots with pre-set responses. There was no flexibility in what I could talk about with the chatbot." This lack of conversational flexibility left some participants feeling disconnected and constrained.

Interestingly, while pure LLM chatbots have been able to generate contextually relevant responses, participants noted that those aligned with expert-crafted dialogues provided more structured, professional, supportive and motivational conversations. These chatbots asked thoughtful questions, provided relevant examples, and offered advice in a structured, encouraging tone. One participant shared, “I liked the way it asked questions, gave examples and advices. I found that they were very relatable. (LLM-SAG (Prompt))”.
This well-balanced flexibility and therapeutic structures for the engaging and therapeutically effective psychotherapy for behavioral interventions. 
However, some participants felt that overly professional or friendly tones could undermine the seriousness of certain topics. “I did not really like the super friend-like way of talking; I prefer a more professional type of way. (LLM-SAG (FT))” one participant noted. This points to the need for chatbots to dynamically adjust their conversational styles based on context, avoiding extremes and tailoring communication to user expectations and emotional needs.

In sum, the conversational styles of chatbots, particularly how they balance flexibility with structure, and friendliness with professionalism, directly influence user-perceived comfort, relatability, and engagement with the chatbots.



\subsubsection{\textbf{\revise{LLM-Powered Intelligence for Engaging Psychotherapy: Personalization, Memory, and Empowerment}}}
\label{qualititive:personalization}

This theme highlights the vital role of LLM-powered intelligence for enhancing digital psychotherapy for behavioral intervention, including features such as personalization, chatbot's memory, and user empowerment in creating engaging and effective chatbot-delivered interventions. 
The rule-based chatbot was disliked because of its pre-set responses and fixed dialogue structures without personalization, as noted, "I enjoy receiving the intended lesson and information in a very straightforward manner. [...] However, it did feel less personal and tailored. It was more like a good lesson than a conversation and coaching."
Most participants appreciated chatbots that responded in a personalized but still structured way, offering tailored advice rather than generic responses. One participant remarked, "It guides my behavior change step by step in human-like conversations. And it is not overwhelming. It can understand my specific questions and give correct answers, and then it can go back to the previous talk after my questions, interesting. (LLM-SAG (Prompt))" 
This kind of personalized guidance, delivered at a comfortable pace, helped participants feel heard, supported, and empowered in their behavioral change journey.

In addition, participants also emphasized that chatbots capable of remembering prior interactions and adapting their responses accordingly made the conversations more coherent, supportive, and relatable. 
One participant shared, "The initial advice was also very generic and not useful for me specifically but in the rest of the conversation, the bot switched to suggestions, which were much more tailored to me and actually did inspire me. I even made a note and am going to implement one of the suggestions. (pure-LLM)"
This illustrates how contextual awareness can strengthen engagement and empower the effectiveness of psychotherapeutic interventions. This ability to build on past interactions can also enhance the continuity and personalization of long-term conversations.

Therefore, these LLM-powered intelligent features could help build more engaging, empowering, and effective chatbot-delivered interventions. By integrating these features and maintaining the right balance of conversational styles, chatbots can better support users’ therapeutic goals and foster more engaging experiences.

\subsubsection{\textbf{\revise{User-Informed Suggestions for Enhancing Chatbot-Delivered Psychotherapy}}}

Participants shared valuable suggestions to improve chatbot-delivered psychotherapy for health behavior change, emphasizing the need for more engaging, adaptable, and therapeutically effective interactions.

A key suggestion was to integrate LLM capabilities into rule-based systems. 
Participants believed this would combine the structured guidance of scripted dialogues with the conversational flexibility, pointing to the potential for more dynamic and personalized interactions. Alongside this, participants highlighted the value of open-ended questions, which LLMs are well-suited to generate. Closed-question formats were seen as limiting, while open dialogue can encourages reflection and deeper engagement in the therapeutic process.

Participants also stressed the importance of modulating the chatbot’s conversational style. A balanced tone, friendly yet professional, was viewed as critical. One participant shared, “It could have used a slightly different tone to balance the formality that enhances its reliability and friendliness.” 

Personalization emerged as another vital area for improvement. Participants wanted chatbots to tailor responses to their specific circumstances and offer actionable advice. One participant noted, “It should have given clearer instructions for users for their change” highlighting the need for more individualized support. In line with this, participants emphasized the importance of memory and contextual awareness. The ability of a chatbot to recall past interactions was seen as essential for maintaining continuity, deepening personalization, and improving therapeutic relevance. 

Collectively, these user insights point toward a future of chatbot-delivered therapy that is more personalized, flexible, engaging, and therapeutically effective, ultimately leading to more impactful psychotherapeutic interventions.

%% file: tables/analysis_results_automatic.tex
\begin{table}[!ht]
\centering
\renewcommand{\arraystretch}{0.90}
\small
\setlength{\tabcolsep}{20.0pt} 
\begin{tabular}{lccc}
\hline
\multirow{1}{*}{\textbf{Chatbot Types}} & \multirow{1}{*}{\textbf{Auto-Metric 1}} & \multicolumn{1}{c}{\textbf{Auto-Metric 2}} \\ \hline

Rule-based (as oracle)
 & 100.00 & 98.81 \\  
\multirow{1}{*}{Pure LLM} 
 & 85.71 & 12.62 \\ 
\multirow{1}{*}{LLM-SAG (FT)} 
 & 71.43 & 70.24 \\ 
\multirow{1}{*}{LLM-SAG (Prompt)} 
 & 97.62 & 96.42 \\ 
\hline

\end{tabular}
\vspace{0.6mm}
\caption{Results (ratio) of automatic metrics for evaluating rule-based, pure LLM, and LLM-SAG chatbots.}
\label{tab:res_automatic}
\vspace{-4.2mm}
\end{table}

%% file: tables/study_1_gee_rule_vs_others.tex
\begin{table*}[!htbp]
\centering
\renewcommand{\arraystretch}{1.0}
\footnotesize
\setlength{\tabcolsep}{2pt}
\begin{tabular*}{\textwidth}{@{\extracolsep{\fill}}p{3.0cm}p{2.6cm}p{3.0cm}p{1.5cm}p{2.1cm}p{1.8cm}@{}}
\hline
\textbf{Comparison} & \textbf{Measure} & \textbf{Mean (SD)} & \textbf{Coefficient} & \textbf{Effect ($Std.B$)} & \textbf{p-value} \\
\hline

\multicolumn{6}{l}{\textbf{Rule-based vs. Pure LLM}} \\
\cline{1-6}
 & Linguistic Quality & 3.66 (0.22) vs. 3.76 (0.14) & 0.10 & 0.14 (small) & 0.45 \\
 & Dialogue Relevance & 3.26 (0.88) vs. 3.09 (0.33) & -0.13 & 0.22 (medium) & 0.34 \\
 & Empathy & 2.53 (0.10) vs. 2.99 (0.18) & 0.45 & 0.44 (medium) & 0.04 $*$ \\
 & Engagement & 2.58 (0.53) vs. 2.72 (0.42) & 0.14 & 0.18 (small) & 0.33 \\
 & Perceived MI & 3.13 (0.15) vs. 3.13 (0.08) & 0.00 & 0.00 (small) & 1.00 \\
 & Motivation Change & 2.86 (0.00) vs. 3.00 (0.00) & 0.14 & 0.13 (small) & 0.47 \\
 & Therapeutic Alliance & 3.31 (0.64) vs. 3.33 (0.36) & 0.02 & 0.02 (small) & 0.92 \\
 & Usability & 3.60 (0.40) vs. 3.56 (0.24) & -0.04 & 0.05 (small) & 0.82 \\
\hline

\multicolumn{6}{l}{\textbf{Rule-based vs. LLM-SAG (FT)}} \\
\cline{1-6}
 & Linguistic Quality & 3.66 (0.22) vs. 3.60 (0.18) & -0.06 & 0.08 (small) & 0.65 \\
 & Dialogue Relevance & 3.26 (0.88) vs. 3.38 (0.51) & 0.15 & 0.25 (medium) & 0.21 \\
 & Empathy & 2.53 (0.10) vs. 2.94 (0.15) & 0.41 & 0.40 (medium) & 0.04 $*$ \\
 & Engagement & 2.58 (0.53) vs. 2.58 (0.54) & 0.00 & 0.00 (small) & 1.00 \\
 & Perceived MI & 3.13 (0.15) vs. 3.31 (0.12) & 0.19 & 0.17 (small) & 0.41 \\
 & Motivation Change & 2.86 (0.00) vs. 3.00 (0.00) & 0.14 & 0.13 (small) & 0.46 \\
 & Therapeutic Alliance & 3.31 (0.64) vs. 3.33 (0.48) & 0.01 & 0.01 (small) & 0.94 \\
 & Usability & 3.60 (0.40) vs. 3.61 (0.26) & 0.01 & 0.01 (small) & 0.97 \\
\hline

\multicolumn{6}{l}{\textbf{Rule-based vs. LLM-SAG (Prompt)}} \\
\cline{1-6}
 & Linguistic Quality & 3.66 (0.22) vs. 3.95 (0.23) & 0.30 & 0.40 (medium) & 0.02 $*$ \\
 & Dialogue Relevance & 3.26 (0.88) vs. 3.39 (0.68) & 0.16 & 0.27 (medium) & 0.20 \\
 & Empathy & 2.53 (0.10) vs. 3.07 (0.16) & 0.53 & 0.52 (large) & 0.01 $**$ \\
 & Engagement & 2.58 (0.53) vs. 2.90 (0.33) & 0.32 & 0.40 (medium) & 0.03 $*$ \\
 & Perceived MI & 3.13 (0.15) vs. 3.74 (0.13) & 0.62 & 0.57 (large) & 0.01 $**$ \\
 & Motivation Change & 2.86 (0.00) vs. 3.42 (0.00) & 0.56 & 0.51 (large) & 0.01 $**$ \\
 & Therapeutic Alliance & 3.31 (0.64) vs. 3.58 (0.46) & 0.26 & 0.32 (medium) & 0.10 \\
 & Usability & 3.60 (0.40) vs. 3.82 (0.26) & 0.22 & 0.30 (medium) & 0.19 \\
\hline

\end{tabular*}
\caption{Results of generalized estimating equations (GEE)~\cite{gee} comparing rule-based with three types of LLM-powered chatbots (pure LLM, LLM-SAG via prompting and fine-tuning). (**$p$<.01, *$p$<.05, \minor{``Coefficient'' represents the unstandardized regression coefficient.})}
\label{tab:study_1_gee_rule_vs_others}
\vspace{-3.60mm}
\end{table*}

%% file: tables/study_1_wilcon.tex
\begin{table*}[!htbp]
\centering
\renewcommand{\arraystretch}{1.1}
\footnotesize
\setlength{\tabcolsep}{2pt}
\begin{tabular*}{\textwidth}{@{\extracolsep{\fill}}p{3.5cm}p{2.8cm}p{3.4cm}p{2.2cm}p{2.0cm}@{}}
\hline
\textbf{Comparison} & \textbf{Measure} & \textbf{Mean (SD)} & \textbf{Effect ($Std.B$)} & \textbf{p-value} \\
\hline

\multicolumn{5}{l}{\textbf{LLM-SAG (Prompt) vs. LLM-SAG (FT)}} \\
\cline{1-5}
 & Linguistic Quality & 3.95 (0.23) vs. 3.60 (0.18) & 0.43 (large) & 0.01 $**$ \\
 & Dialogue Relevance & 3.39 (0.68) vs. 3.38 (0.51) & 0.09 (small) & 0.54 \\
 & Empathy & 3.07 (0.16) vs. 2.94 (0.15) & 0.07 (small) & 0.67 \\
 & Engagement & 2.90 (0.33) vs. 2.58 (0.54) & 0.36 (medium) & 0.02 $*$ \\
 & Perceived MI & 3.74 (0.13) vs. 3.31 (0.12) & 0.40 (medium) & 0.01 $**$ \\
 & Motivation Change & 3.42 (0.00) vs. 3.00 (0.00) & 0.35 (medium) & 0.02 $*$ \\
 & Therapeutic Alliance & 3.58 (0.46) vs. 3.33 (0.48) & 0.27 (medium) & 0.08 \\
 & Usability & 3.82 (0.26) vs. 3.61 (0.26) & 0.24 (medium) & 0.11 \\
\hline
\end{tabular*}
\vspace{-0mm}
\caption{Results from Wilcoxon signed-rank test~\cite{wilcoxon} for pairwisely comparing the LLM-powered chatbots aligned with the full expert-crafted dialogue scripts (SAG) through prompting or fine-tuning. (**$p$<.01, *$p$<.05)}
\label{tab:analysis_results_2}
\vspace{-2.0mm}
\end{table*}

%% file: 6-Study-2-Methods.tex
\section{Study 2: A Collaborative and Scalable Approach to Develop LLM-Powered Psychotherapy Chatbots Aligned with Partial Expert-Crafted Dialogues}

Study 1 underscored the value of expert-crafted dialogue scripts for aligning LLMs in psychotherapy.
However, as shown in Fig~\ref{fig:ssag} and Fig~\ref{fig:dataset_example}, \revise{the most labor-intensive aspects of the expert-scripting involve both creating dialogue content (e.g., questions, reflections, advice) and designing structured dialogue flows. SAG relies heavily on fully expert-authored scripts, which are costly and difficult to scale in expert-driven domains like psychotherapy.}

To address this, we introduced Script-Strategy Aligned Generation (SSAG), a more flexible and efficient alignment approach that \revise{requires only partial expert input.} 
SSAG drew on evidence-based therapeutic strategies~\cite{Therapeutic_Skills_Affect_Clinical, MI_Strategies_cscw} adapted from Motivational Interviewing (MI)~\cite{cos_xsun, misc-1} to guide LLMs in generating responses that balance therapeutic adherence with conversational flexibility.
\revise{SSAG enabled LLMs to co-author therapeutic dialogue content and dynamically manage dialogue flow based on predicted therapeutic strategies, reducing reliance on \minor{experts' fully scripted dialogue.}}
In Study 2, \revise{we compared SSAG (partial alignment) with LLM-SAG (full alignment) to assess whether SSAG could maintain therapeutic effectiveness while reducing dependence on expert-authored dialogue scripts \minor{(RQ2 and RQ3)}}. \minor{By streamlining expert input demands and enabling flexible generation, SSAG facilitated more efficient and scalable development of psychotherapy chatbots supported by LLMs.}


\subsection{Approach: Script-Strategy Aligned Generation (SSAG)}
\label{ssag_approach}

Script-Strategy Aligned Generation (SSAG) combined \revise{partial expert-scripted dialogue content} with therapeutic strategies to \minor{reduce reliance on fully expert-authored dialogue scripts}. 
As shown in Fig~\ref{fig:ssag} and Fig~\ref{fig:dataset_example}, expert-crafted dialogue scripts typically include \revise{five components: psychotherapeutic topics, therapeutic questions, reflections, and general advice, along with tree-structured dialogue flows.
SSAG simplified this by requiring only the three most essential components: psychotherapeutic topics, therapeutic questions, and general advice (optional), as expert pre-authored input.}
\minor{Unlike prior alignment approaches such as unconstrained prompting/fine-tuning that rely heavily on full scripting and limit the therapeutic explainability, SSAG introduced a middle-ground alignment grounded in strategy prediction, 
offering greater modularity, efficiency, and explainability than direct prompting or fine-tuning alone.}
Specifically, SSAG operated in the following two steps as illustrated in Fig~\ref{fig:ssag}:

\textbf{Step 1: Predicting the therapeutic strategy:}
SSAG first predicts the next therapeutic strategy based on dialogue context, selecting from core therapeutic behaviors: "asking questions," "reflective listening," and "giving advice."

\textbf{Step 2: Generating the response:}
Once the therapeutic strategies are selected, LLM generates a response according to the predicted strategies. For example, if "asking questions" or "giving advice" is selected, the LLM retrieves and paraphrases the pre-written expert content (therapeutic questions or advice) based on dialogue context. 
If "reflective listening" is selected, the LLM generates reflections freely based on the context, allowing for dynamic, empathetic responses without scripting.
By anchoring responses to specific therapeutic strategies, SSAG aligns the conversation with therapeutic principles~\cite{MI_Strategies_cscw,Therapeutic_Skills_Affect_Clinical}, maintaining control over the dialogue flow while enabling flexibility in generation.



\subsection{Study Methods}

\subsubsection{Study design and procedure}

\camera{The procedure of Study 2 is illustrated in Fig~\ref{fig:procedure} (b).}
We conducted a 10-day field study employing a mixed design to compare three chatbot types: (1) a rule-based chatbot (baseline), (2) an LLM-powered chatbot aligned via SAG using prompting (LLM-SAG (Prompt)), and (3) an LLM-powered chatbot aligned via the flexible SSAG approach (LLM-SSAG). 
Participants were divided into two groups: one interacted with the rule-based chatbot and LLM-SAG (Prompt); the other with the rule-based chatbot and LLM-SSAG.
Each group followed a counterbalanced design, where participants used one chatbot for five consecutive days, then switched to the other, totaling ten days. 
After each daily interaction, participants completed evaluation to assess their experience with the chatbot. 
This design enabled direct comparisons under consistent conditions \revise{to evaluate whether SSAG (partial alignment) could deliver psychotherapeutic interventions as effectively as SAG (full alignment).}


\subsubsection{Types of chatbots}
Study 2 examined the following three types of chatbot: 

\noindent 1) \textbf{rule-based chatbot.}
\revise{This chatbot was implemented using the RASA framework~\cite{rasa}, as same as the rule-based chatbot used in Study 1 (see Section~\ref{study_1_chatbots} (1)). 
}

\noindent 2) \textbf{LLM-SAG (Prompt).}
This chatbot was implemented by aligning with the \revise{full expert-crafted dialogue scripts using SAG via prompting} as in Study 1 (see Section~\ref{study_1_chatbots} (4))

\noindent 3) \textbf{LLM-SSAG.}
\revise{
As detailed in Section~\ref{ssag_approach}, SSAG offered a flexible alternative to SAG by reducing the need for fully scripted dialogues. It relied on three core expert-crafted elements: psychotherapeutic topics, a sequence of therapeutic questions, and general advice (optional).}
LLM-SSAG chatbot operated in two steps (see~Fig~\ref{fig:ssag}). In Step 1, LLM-1 predicted the therapeutic strategies (i.e., asking questions, reflective listening, or giving advice) based on dialogue context. In Step 2, LLM-2 generated the dialogue response, ensuring alignment with the predicted strategies from Step 1.


In this study, \revise{SSAG restricted completely ``free'' generation to reflective responses, while therapeutic questions and advice were retrieved and paraphrased from expert-crafted dialogues. This balanced flexibility and safety: reflections are lower-risk and promote empathy and engagement, while questions and advice, being more directive and sensitive, are anchored in validated expert input to maintain therapeutic integrity and ethical standards.
However, SSAG was designed to be extensible, similar to Retrieval-Augmented Generation (RAG)~\cite{rag}, but with greater controllability over which response types can be freely generated and which must remain grounded in expert data.} 
Although this work restricted question and advice generation anchoring to expert-authored content for safety, \revise{future versions could allow more personalized generations with additional validation mechanisms to ensure safety and adherence to ethical standards.}
This setting also enhanced efficiency and adaptability. \revise{Training LLMs to freely generate expert-quality therapeutic questions and advice would require extensive domain-specific data and frequent model retraining.}
Findings from Study 1 have evidenced that prompting provides a more efficient alignment for expertise integration than fine-tuning. 
By contrast, \revise{SSAG maintained alignment with expert guidance while enabling flexible generation and dynamic dialogue flow management, reducing the need for continuous model retraining and making it more scalable.}

\revise{The prompt templates used for the LLM-powered chatbots are provided in~\ref{appendix:prompt-template-2} and ~\ref{appendix:prompt-template-3}}


\subsubsection{Self-developed application}

\begin{figure*}[!htbp]
\centering
\includegraphics[width=0.960\textwidth]{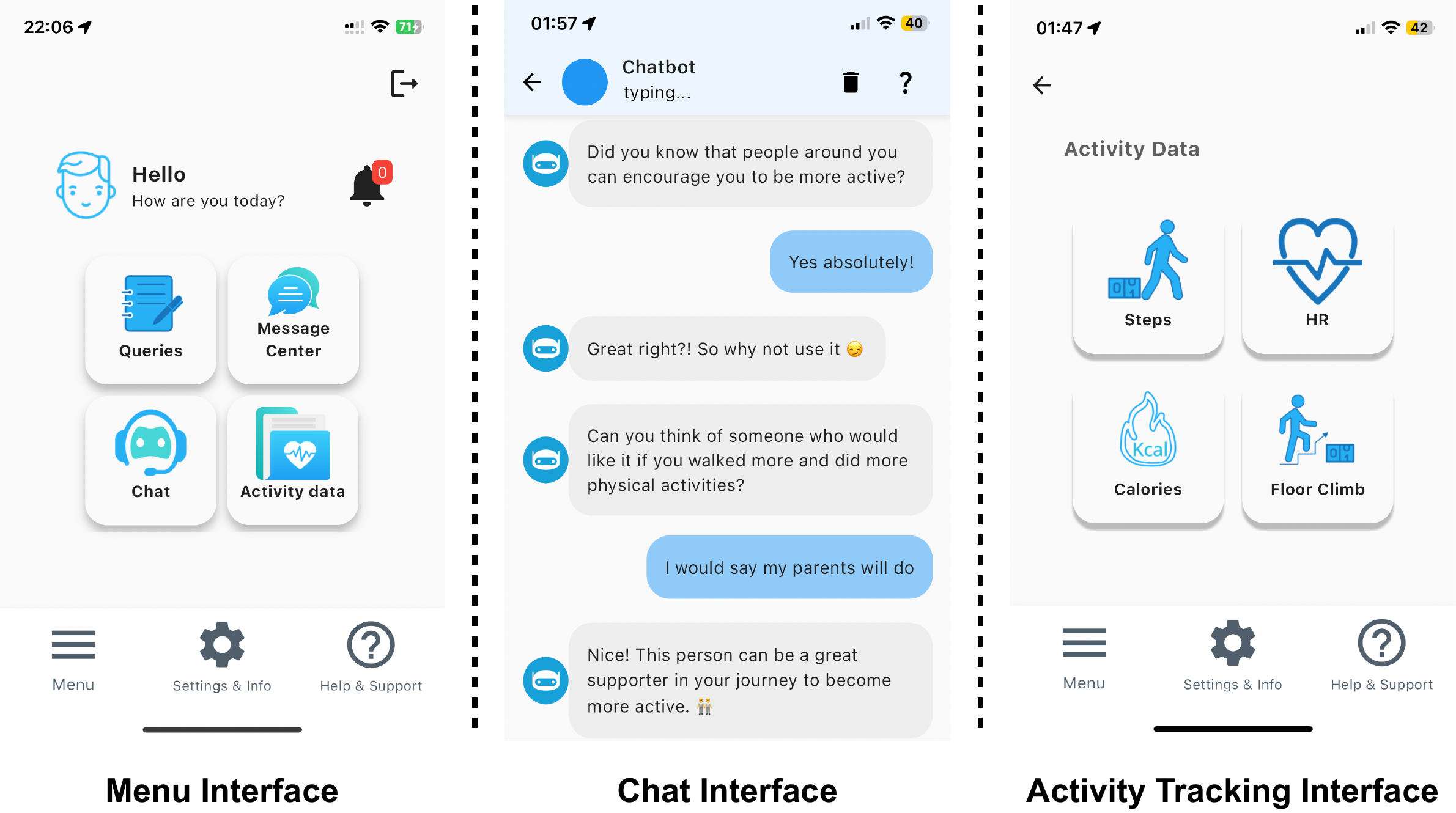}
\caption{A mobile application was developed for a field study to enable long-term chatbot interactions and the tracking of PA levels.}
\label{fig:study_2_interface}
\end{figure*}

We developed a mobile app for Study 2 (see Fig~\ref{fig:study_2_interface}) to support chatbot interaction and physical activity tracking. 
The app included: (1) a Menu Interface for navigation; (2) a Chat Interface for psychotherapy sessions, and (3) an Activity Tracking Interface to monitor daily steps and walking distance. 
This integrated design of the app enabled a seamlessly interactive experience for chatbot interventions and evaluations.

\subsubsection{Measures}

To ensure consistency with Study 1, we used the same evaluation metrics, including linguistic quality, engagement, perceived empathy and MI level, therapeutic alliance, and usability. These were assessed daily after each chatbot interaction. 
Intrinsic motivation to change was measured every two days (pre-, Day 2, Day 4, and post-intervention) using a validated scale adapted from~\cite{motivation_measure}. 
Physical activity (i.e., daily steps and distance) was continuously tracked throughout the study to evaluate the effectiveness of physical activity promotion.


\subsubsection{Participants}

Based on a G*Power analysis~\cite{gpower}, a minimum of 16 participants was required to detect a medium effect size ($d$ = 0.25) with 80\% power and $\alpha = 0.05$. 
To ensure robustness, 21 participants were recruited (N = 21). All participants were 18 or older, fluent in English, and recruited through institutional channels and social media. 
They received monetary compensation aligned with the guidelines for completing the participation over 10 days. The study was approved by the institutional ethics committee. 
Participant demographics are shown in Table~\ref{table:demographics_study_2}.

\vspace{-1.6mm}

\input{tables/demographics_study_2}



\subsubsection{Data analysis}

We first checked data suitability using the Shapiro-Wilk test~\cite{SHAPIRO1965} for normality and Bartlett’s test~\cite{Arsham2011} for homogeneity of variance. 
As data were not fully normally distributed, we applied Generalized Estimating Equations (GEE)~\cite{gee}, a statistical method robust to non-normality, to compare participants' perception and therapeutic effectiveness across rule-based, LLM-SAG (Prompt), and LLM-SSAG chatbot conditions.

%% file: tables/demographics_study_2.tex
\begin{table}[!h]
\centering
\footnotesize
\renewcommand{\arraystretch}{1.0}
\begin{tabularx}{\columnwidth}{p{4cm} >{\raggedright\arraybackslash}p{7cm} >{\raggedright\arraybackslash}X}
\toprule
\textbf{Demographic} & \textbf{Categories} & \textbf{Numbers of Participants (\%)} \\
\midrule
Gender & Female & 13 (61.9\%)
\\ 
 & Male & 8 (38.1\%)
\\
\hline
Age & 18-24 & 9 (42.9\%)
\\
 & 25-34 & 10 (47.6\%)
\\
 & 35-44 & 1 (4.8\%)
\\
 & 45-54 & 1 (4.8\%)
\\
\hline
Education & High school degree or equivalent & 2 (9.5\%)
\\
 & Bachelor’s degree & 5 (23.8\%)
\\
 & Master’s degree & 10 (47.6\%)
\\
 & Doctorate or higher & 4 (19.0\%)
\\
\hline
Professional Domain & Health and Medical Science & 3 (14.3\%)
\\
 & Science, Technology, Engineering, Mathematics (STEM) & 5 (23.8\%)
\\
 & Business, Economics, and Law & 3 (14.3\%)
\\
 & Communication, Arts, Culture and Entertainment & 4 (19.0\%)
\\
 & Education and Social Science & 3 (14.3\%)
\\
 & Government and Public Sector & 2 (9.5\%)
\\
& Other & 1 (4.8\%)
\\
\bottomrule
\end{tabularx}
\caption{Characteristics of participants in Study 2.}
\vspace{-4.0mm}
\label{table:demographics_study_2}
\end{table}

%% file: 6-Study-2-Results.tex
\subsection{Findings}

\subsubsection{Pre-validation of therapeutic strategy prediction in SSAG}

\input{tables/study_2_ssag_step_1}

To support the feasibility of SSAG, we first validated LLMs' ability to predict therapeutic strategies (Step 1 of SSAG as defined in Section~\ref{ssag_approach}), which serve as the basis for response generation in Step 2 of SSAG. 
Building on prior work~\cite{cos_xsun,mi_pred_1}, we benchmarked LLMs using two open-sourced MI datasets: AnnoMI~\cite{annomi_1} and BiMISC~\cite{bimisc} for evaluating this prediction task.

As shown in Table~\ref{tab:automatic-cos}, GPT-4 (zero-shot setting) achieved strong performance, while fine-tuning on data further improved accuracy (GPT-4o (FT)). Notably, performance was lower on the BiMISC dataset due to its multi-strategy complexity (i.e., one utterance has multiple MI strategies), compared to the single-strategy in the AnnoMI dataset. 
\camera{Although the overall accuracy was modest, this is acceptable within SSAG, where multiple strategies may be contextually appropriate and no single strategy is universally optimal.}
These results confirmed that LLMs can effectively support SSAG’s Step 1 for therapeutic strategy prediction and serve as a basis for guiding dialogue generation in Step 2 of SSAG.


\subsubsection{\revise{Comparing chatbot types: rule-based vs. LLM-SAG vs. LLM-SSAG}}
\label{finding_rq2}

\input{tables/study_2_gee_rule_vs_others}

We evaluated three chatbot types: rule-based, LLM-SAG (Prompt), and LLM-SSAG in Study 2 using consistent measures from Study 1.
As shown in Table~\ref{tab:study_2_gee_rule_vs_others}, 
\revise{LLM-powered chatbots aligned with partial expert-crafted scripts (LLM-SSAG) performed comparably to those aligned with full scripts (LLM-SAG) in both conversational quality and therapeutic effectiveness. Besides, both of LLM-powered chatbots significantly outperformed rule-based chatbots in delivering psychotherapy across various assessing dimensions.}





    



\begin{figure*}[!htbp]
\centering
\includegraphics[width=0.576\textwidth]{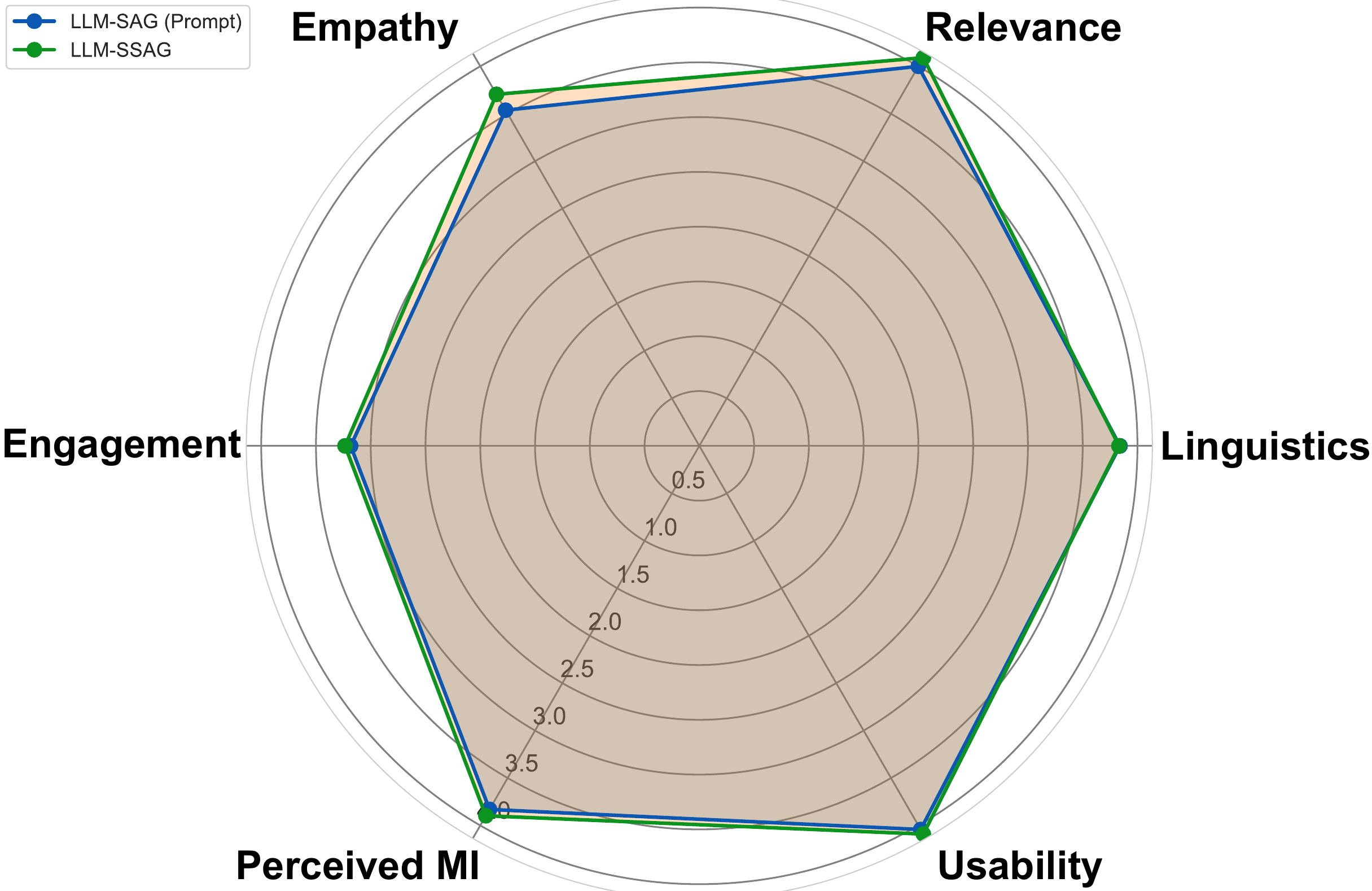}
\caption{Comparison (mean value) between the LLM-powered chatbots employing strict SAG through prompting (LLM-SAG (Prompt)) vs. our proposed flexible alignment (LLM-SSAG).}
\vspace{-2.0mm}
\label{study_2_radar}
\end{figure*}

\revise{To address RQ2, we compared LLM-SAG (Prompt) and LLM-SSAG to evaluate whether LLM-SSAG chatbot could achieve comparable therapeutic effectiveness and conversational quality to those aligned with full pre-scripted dialogues by SAG.}
Results from the GEE analysis (Table~\ref{tab:study_2_gee_rule_vs_others}) showed no significant differences in performance between LLM-SSAG and LLM-SAG across key metrics, including linguistic quality (p=0.97), dialogue relevance (p=0.58), empathy (p=0.63), engagement (p=0.89), MI adherence (p=0.73), therapeutic alliance (p=0.73), and usability (p=0.86). 
\revise{The findings that no significant performance differences exist between LLM-SSAG and LLM-SAG across most metrics supported the effectiveness of the partial alignment approach in LLM-SSAG, maintaining similar levels of conversational quality and therapeutic adherence while offering a more flexible alternative to the fully aligned LLM-SAG chatbots for psychotherapy.
Radar plots in Fig~\ref{study_2_radar} explicitly visualized their comparable performances.}


\begin{figure*}[!ht]
    \centering
    \begin{subfigure}[t]{0.986\textwidth}
        \centering
        \includegraphics[width=\textwidth]{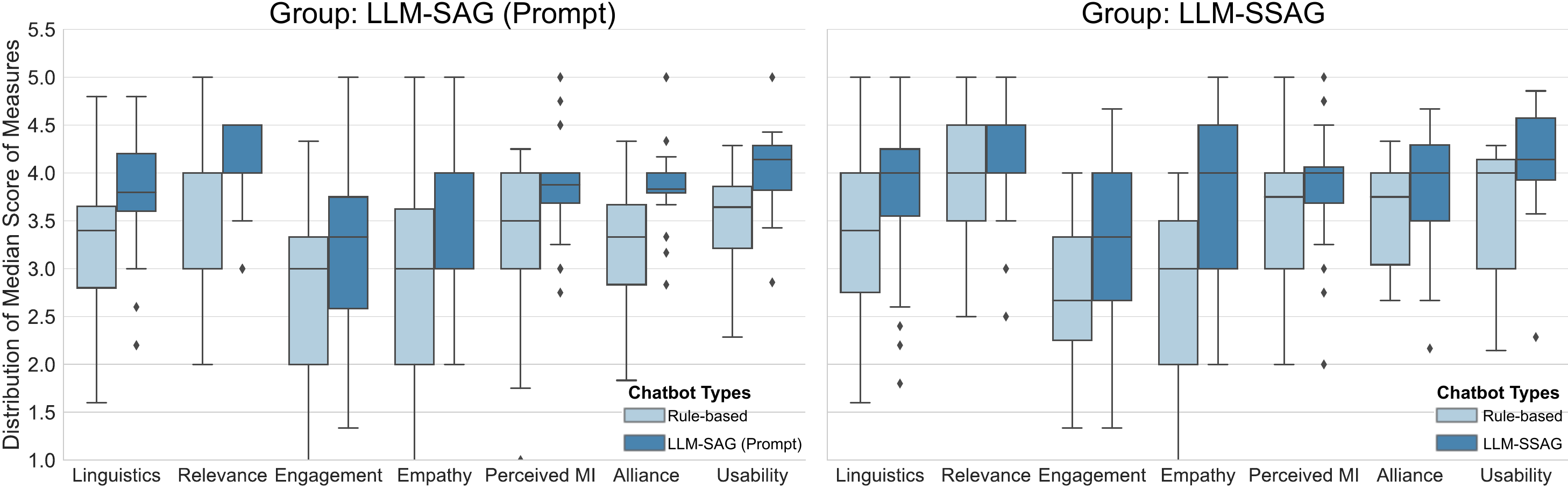}
        \vspace{-4mm}
        \caption{Medians of measures in human evaluation across chatbot conditions.}
        \label{fig:study_2_gee_analysis}
    \end{subfigure}

    \vspace{2mm} 

    \begin{subfigure}[t]{0.986\textwidth}
        \centering
        \includegraphics[width=\textwidth]{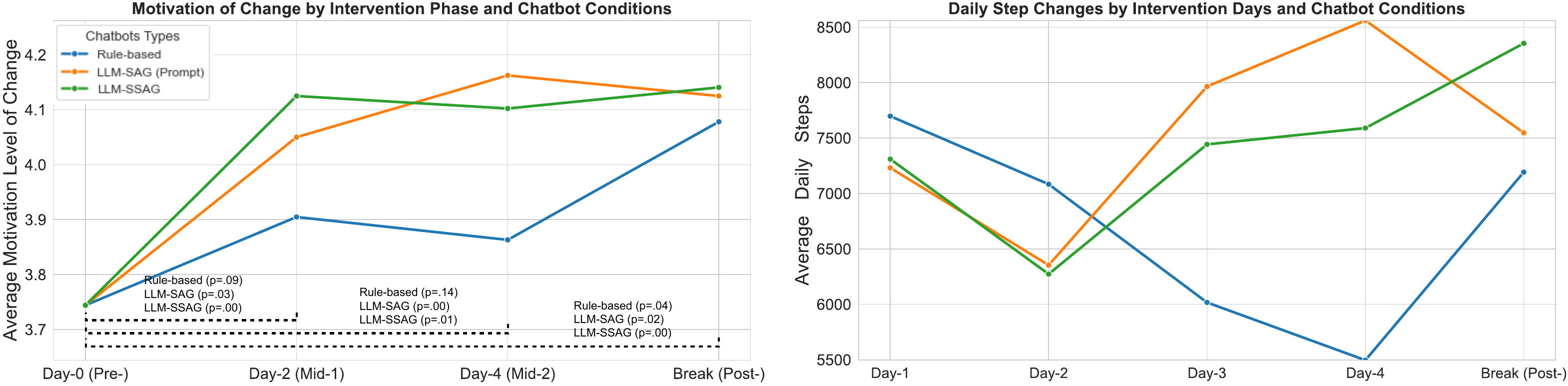}
        \vspace{-4mm}
        \caption{Means of motivation change and the daily steps during the study.}
        \label{fig:gee_analysis_motivation}
    \end{subfigure}
    \vspace{-3mm}    
    \caption{Analysis results of chatbot evaluation, motivation change, and daily steps across intervention days and chatbot conditions.}
    \label{fig:motivation_activity}

    \vspace{-3.0mm} 
\end{figure*}

As shown in Table~\ref{tab:study_2_gee_rule_vs_others} and Fig~\ref{fig:motivation_activity} (a), linguistic quality (LLM-SSAG: M=3.83; LLM-SAG: M=3.84) and dialogue relevance (LLM-SSAG: M=4.09; LLM-SAG: M=4.00) of both LLM-powered chatbots were rated higher than the rule-based chatbot baseline, indicating better conversational quality. Empathy (LLM-SSAG: M=3.70; LLM-SAG: M=3.54), engagement (LLM-SSAG: M=3.23; LLM-SAG: M=3.18), MI adherence (LLM-SSAG: M=3.90; LLM-SAG: M=3.83), and usability (LLM-SSAG: M=4.09; LLM-SAG: M=4.04) were also higher than the rule-based baseline, with LLM-SSAG slightly outperforming LLM-SAG. Both LLM-powered chatbots scored higher in therapeutic alliance (LLM-SSAG: M=3.83; LLM-SAG: M=3.87) than rule-based chatbots, reflecting stronger therapeutic rapport.

In terms of the actual behavioral outcomes, as shown in Fig~\ref{fig:motivation_activity} (b), both LLM-SAG and LLM-SSAG chatbots showed sustained increases in motivation change and daily step counts throughout the intervention period, while the rule-based chatbot showed limited impact on motivation change over time. 
As for the daily step counts, while all conditions started with similar step counts on Day 1, the trajectories diverged significantly. 
LLM-SAG maintained a steady increase, peaking on Day 4. LLM-SSAG followed a similar upward trend, showing consistent increases and achieving the highest average steps by the end of the study.
In contrast, the rule-based chatbot exhibited a downward trend, reaching its lowest point on Day 4 before a slight recovery. 
These results 
\revise{demonstrated that LLM-SSAG chatbot can offer comparable performance to fully aligned LLM-SAG chatbot.}


\subsubsection{\revise{Comparison of reliance on expert-crafted content: full alignment (SAG) vs. partial alignment (SSAG)}}
\label{finding_rq3}

\revise{To address RQ3,} SSAG supported efficient co-development where experts pre-script only essential dialogue elements (i.e., psychotherapeutic topics, key therapeutic questions, health advice (optional)), and LLMs handle response generation or paraphrase as well as dynamic dialogue flow management based on therapeutic strategies and dialogue context, \minor{which reduces the need for fully expert-scripted content and structured flows, thereby could lower expert authoring workload.}

\revise{To quantify this reduction, we compared the number of pre-authored utterances required for SSAG versus SAG. The LLM-SAG (Prompt) chatbot utilized the full expert-crafted dialogue scripts, consisting of 1,876 utterances and complete tree-structured flows (see Fig~\ref{fig:dataset_example}). 
In contrast, LLM-SSAG operated with only 501 therapeutic questions and 203 general advice utterances, covering only 37.5\% of the total utterances in the dataset. 
A detailed breakdown by psychotherapeutic topic is included in~\ref{appendix_dialogue}. 
These results showed that SSAG requires less than two-fifths of the expert-authored dialogue content needed for SAG.
}\minor{While this suggested a potential reduction in expert authoring workload, we did not directly measure time or workload. Thus, these findings should be interpreted as indirect evidence of efficiency, not a conclusive workload evaluation.}


%% file: tables/study_2_ssag_step_1.tex
\begin{table}[!ht]
\renewcommand{\arraystretch}{0.70} 
\small
\setlength\tabcolsep{8.0pt}
\centering
\begin{tabular}{lcccccccc}
\toprule

Dataset & \multicolumn{4}{c}{AnnoMI} & \multicolumn{4}{c}{BiMISC} \\ 

\cmidrule(lr){2-5} \cmidrule(lr){6-9}

Metric   &     & Accuracy    & F1    &    &     & Accuracy    & F1   & \\ 

\midrule
Flan-T5~\cite{flan-t5} &  & 46.2 & 77.6 &  &  & 19.1 & 17.4 \\

Vicuna-13B~\cite{vicuna} &  & 44.7 & 76.2 &  &  & 10.5 & 18.8 \\

GPT-4 &  & \textbf{50.0} & \textbf{78.9} &  &  & \textbf{33.6} & \textbf{27.9} \\

GPT-4o (FT) &  & \textbf{63.6} & \textbf{81.4} &  &  & \textbf{47.2} & \textbf{36.9} \\

\bottomrule
\end{tabular}
\vspace{0mm}
\caption{Therapeutical strategy prediction on dataset AnnoMI (single-strategy prediction) and BiMISC (multiple-strategy prediction).}
\vspace{-3.6mm}
\label{tab:automatic-cos}
\end{table}

%% file: tables/study_2_gee_rule_vs_others.tex
\begin{table*}[!htbp]
\centering
\renewcommand{\arraystretch}{1.16}
\footnotesize
\setlength{\tabcolsep}{2pt}
\begin{tabular*}{\textwidth}{@{\extracolsep{\fill}}p{3.4cm}p{2.4cm}p{2.8cm}p{1.5cm}p{2.1cm}p{1.8cm}@{}}
\hline
\textbf{Comparison} & \textbf{Measure} & \textbf{Mean (SD)} & \textbf{Coefficient} & \textbf{Effect ($Std.B$)} & \textbf{p-value} \\
\hline

\multicolumn{6}{l}{\textbf{Rule-based vs. LLM-SAG (Prompt)}} \\
\cline{1-6}
 & Linguistic Quality & 3.26 (0.79) vs. 3.84 (0.63) & 0.57 & 0.72 (large) & 0.01 $**$ \\
 & Dialogue Relevance & 3.75 (0.70) vs. 4.00 (0.42) & 0.22 & 0.36 (medium) & 0.18 \\
 & Empathy & 2.81 (1.13) vs. 3.54 (0.84) & 0.75 & 0.70 (large) & 0.00 $**$ \\
 & Engagement & 2.69 (0.89) vs. 3.18 (0.98) & 0.48 & 0.54 (large) & 0.07 \\
 & Perceived MI & 3.38 (0.74) vs. 3.83 (0.52) & 0.41 & 0.62 (large) & 0.04 $*$ \\
 & Therapeutic Alliance & 3.28 (0.65) vs. 3.87 (0.52) & 0.49 & 0.62 (large) & 0.02 $*$ \\
 & Usability & 3.50 (0.60) vs. 4.04 (0.59) & 0.51 & 0.71 (large) & 0.03 $*$ \\
\hline

\multicolumn{6}{l}{\textbf{Rule-based vs. LLM-SSAG}} \\
\cline{1-6}
 & Linguistic Quality & 3.35 (0.83) vs. 3.83 (0.73) & 0.50 & 0.63 (large) & 0.06 $-$ \\ 
 & Dialogue Relevance & 3.89 (0.74) vs. 3.95 (0.23) & 0.23 & 0.37 (medium) & 0.21 \\
 & Empathy & 2.67 (0.95) vs. 3.70 (1.00) & 1.00 & 0.94 (large) & 0.01 $**$ \\
 & Engagement & 2.74 (0.69) vs. 3.23 (0.93) & 0.50 & 0.56 (large) & 0.10 \\
 & Perceived MI & 3.59 (0.69) vs. 3.90 (0.56) & 0.35 & 0.53 (medium) & 0.09 \\
 & Therapeutic Alliance & 3.56 (0.49) vs. 3.83 (0.63) & 0.37 & 0.53 (large) & 0.10 \\
 & Usability & 3.56 (0.85) vs. 4.09 (0.72) & 0.56 & 0.78 (large) & 0.04 $*$ \\
\hline

\multicolumn{6}{l}{\textbf{LLM-SAG (Prompt) vs. LLM-SSAG}} \\
\cline{1-6}
 & Linguistic Quality & 3.84 (0.63) vs. 3.83 (0.73) & 0.01 & 0.01 (small) & 0.97 \\
 & Dialogue Relevance & 4.00 (0.42) vs. 4.09 (0.55) & 0.09 & 0.15 (small) & 0.58 \\
 & Empathy & 3.54 (0.84) vs. 3.70 (1.00) & 0.17 & 0.16 (small) & 0.63 \\
 & Engagement & 3.18 (0.98) vs. 3.23 (0.93) & 0.05 & 0.06 (small) & 0.89 \\
 & Perceived MI & 3.83 (0.52) vs. 3.90 (0.56) & 0.07 & 0.10 (small) & 0.73 \\
 & Therapeutic Alliance & 3.87 (0.52) vs. 3.83 (0.63) & 0.04 & 0.10 (small) & 0.73 \\
 & Usability & 4.04 (0.59) vs. 4.09 (0.72) & 0.05 & 0.07 (small) & 0.86 \\
\hline

\multicolumn{6}{l}{\textbf{Motivation Change Day-1 vs. Day-5}} \\
\cline{1-6}
 & Rule-based & 3.74 (0.70) vs. 4.07 (0.35) & 0.11 & 0.20 (medium) & 0.04 $*$ \\
 & LLM-SAG (Prompt) & 3.74 (0.70) vs. 4.13 (0.33) & 0.39 & 0.67 (large) & 0.00 $**$ \\
 & LLM-SSAG & 3.74 (3.70) vs. 4.14 (0.65) & 0.44 & 0.75 (large) & 0.00 $**$ \\
\hline
\end{tabular*}
\caption{Results of generalized estimating equations (GEE)~\cite{gee} comparing rule-based and LLM-powered chatbots via full alignment (SAG) and partial alignment (SSAG). (**$p$<.01, *$p$<.05, \minor{``Coefficient'' represents the unstandardized regression coefficient.})}
\label{tab:study_2_gee_rule_vs_others}
\vspace{-4mm}
\end{table*}

%% file: 7-Discussion.tex
\section{Discussion}

\subsection{Expert-Crafted Dialogues Remain Essential for Chatbot-Delivered Psychotherapy in the Era of LLMs}

Chatbots for psychotherapy~\cite{linwei-mi,Sun2023VirtualSupport,smoking-chatbot} have traditionally relied on rule-based systems structured around expert-crafted dialogue scripts~\cite{conversation_design_1,conversation_design_2,conversation_design_3}, valued for their safety, controllability, and strict adherence to evidence-based psychotherapeutic techniques like MI and CBT. While reliable, these systems often produce rigid, non-adaptive conversations~\cite{rule-based-system}, limiting their ability to tailor responses to nuanced client needs.
The emergence of LLMs introduces new possibilities for more personalized, empathetic, flexible, and engaging dialogue~\cite{generative_AI_chatbot_mental_health,PELAU2021106855,Empathetic}. 
However, when deployed without domain alignment, pure LLMs, such as ChatGPT~\cite{chatgpt}, lack the domain-specific knowledge, structure, and safety controls needed for high-stakes contexts like psychotherapy. They struggle to generate sequential, goal-directed therapeutic questions and can deviate from evidence-based psychotherapeutic techniques, making experts in such domains hesitate in relying on them alone. 

Study 1 compared four chatbot types: rule-based, pure LLM, and two LLMs aligned with expert-crafted dialogue scripts using fine-tuning or prompting (SAG). Results showed that LLM-aligned chatbots significantly outperformed both rule-based and pure LLMs across assessing metrics. 
These findings \revise{addressed \textbf{RQ 1}} that \textit{expert-crafted dialogue scripts remain essential in the era of LLMs for effective digital psychotherapy.}
Specifically, findings from Study 1 revealed that participants rated LLM-powered chatbots higher than rule-based ones in linguistic quality, empathy, and engagement, highlighting the conversational fluency and personalization that generative models bring. 
While rule-based chatbots are reliable, their rigidity and reliance on pre-authored scripts limit their ability to adapt to diverse conversational contexts~\cite{conversation_design_2}. Participants noted a lack of contextual awareness and personalization, both essential for empathetic and engaging psychotherapy interactions.
\revise{However, the flexibility of LLMs came with tradeoffs.} Pure LLMs were perceived as more natural and engaging but often lacked therapeutic structure and relevance. Participants described them as “human-like” yet vague or unfocused with qualitative insights, resulting in lower scores for dialogue relevance and MI adherence compared to rule-based and aligned LLMs.
In contrast, rule-based chatbots maintained therapeutic objectives and structure but were often seen as rigid and less engaging. 
Aligned LLMs (SAG) effectively bridged this gap by preserving the structure of expert scripts while introducing empathetic and adaptive dialogue generation. Participants appreciated this combination, describing aligned LLM-powered chatbots as “professional yet warm” and capable of delivering structured guidance with emotional resonance, an essential balance in psychotherapy~\cite{therapy_cbt_llm,linwei-mi}.
\revise{The qualitative insights reinforced the quantitative findings: aligned LLMs offered a more effective tradeoff between therapeutic rigor and conversational naturalness than either rule-based or unaligned pure LLMs.}

\revise{Although LLMs are increasingly capable of generating empathetic and engaging dialogue~\cite{ai_show_empathy_1,ai_show_empathy_2}, such AI-generated empathy should be interpreted with caution because 
AI perceives warmth, empathy differently than humans~\cite {Machine_and_Human_Understanding_of_Empathy}. For instance, LLMs may over-express empathy~\cite{navigating_empathy,Therapeutic_Skills_Affect_Clinical}, which can lead to discomfort or unintended emotional effects, and AI often fails to express empathy in positive circumstances while humans do~\cite{navigating_empathy}. A study~\cite{illusion_empathy} shows conversational agents adapt empathetic responses to certain identities, which can be potentially exploitative. 
While our findings show that participants perceived LLM-generated responses as more empathetic and engaging than rule-based ones, such empathy and engagement need further control. 
Prior work suggests that human-AI collaboration can enhance perceived empathy~\cite{Human_AI_collaboration_enables_more_empathic}, supporting our approach of aligning LLMs with expert guidance to balance conversational attributes (e.g., empathy) with therapeutic structure. 
Given the relatively short duration of our study, future research should examine how human-perceived chatbot's empathy evolves over longer-term use.}

In sum, \revise{expert-crafted dialogue scripts remain essential in the era of LLMs, not as rigid blueprints but as scaffolding that enables LLMs to deliver therapeutically-structured, safe, and engaging dialogues.}
SAG-aligned LLMs guided by the expert-crafted scripts outperformed both rule-based and pure LLMs by effectively managing the tradeoff, offering a practical path for engaging and therapeutically effective chatbot-delivered psychotherapeutic interventions.


\subsection{Effective Alignment of LLMs with Domain Expertise for Psychotherapy}


In psychotherapy, aligning LLMs with expert knowledge is essential to ensure conversations remain therapeutically relevant and effective. Prior work on instructing and aligning LLMs~\cite{cos_xsun,therapy_cbt_llm,generative_AI_chatbot_mental_health} has focused primarily on two alignment approaches: fine-tuning and prompting, each with distinct trade-offs. 
Fine-tuning~\cite{fine_tuning} embeds domain knowledge directly into model parameters but is resource-intensive and requires large, curated datasets, often scarce in sensitive fields like psychotherapy. 
In contrast, prompting~\cite{prompt_methods} offers a cost-effective and flexible alternative by guiding LLMs at inference time using structured inputs, avoiding retraining while encoding therapeutic goals through prompt design. 
Prompting has been shown to use less than one-tenth the computational cost of fine-tuning for achieving similar alignment~\cite{when_to_ft}, making it a more scalable solution for psychotherapy where domain-specific data is limited but adherence to therapeutic principles is critical. It also allows quick adaptation to new therapeutic techniques, supporting flexible deployment in evolving psychotherapeutic settings.

\revise{To explore RQ1.2,} Study 1 compared these two alignment approaches for their effectiveness in delivering health interventions using the same expert-crafted dialogue scripts. 
The findings revealed that the prompted chatbot consistently outperformed the one by fine-tuning across assessing measures. 
Prompting preserved the structure of expert-authored dialogue flows while offering greater adaptability in response generation, better supporting context-aware, engaging, and goal-directed therapeutic conversations. 
\revise{This difference in our work partly arises from how each approach handles dialogue structure: fine-tuning embeds the tree structure into the model, reducing adaptability, while prompting uses tree-structured scripts~\cite{tree_of_thoughts} during inference, enabling more flexible, diverse conversational paths while preserving therapeutic coherence and staying on the scripted therapeutic path, as noted by our participants, ``It (LLM-SAG (Prompt)) can understand my specific questions and give correct answers, then it can go back to previous talk after my questions, interesting.''}
\revise{This tradeoff is particularly important in psychotherapy, as prior work~\cite{therapy_cbt_llm,cos} indicates that adherence to therapeutic principles should not come at the expense of empathy and conversational engagement.}

\revise{However, both alignment approaches (i.e., SAG through either prompting and fine-tuning) rely heavily on extensive expert input, making scalability and efficiency a challenge. 
We thereby proposed Script-Strategy Aligned Generation (SSAG), a more flexible alignment that reduces scripting overhead by leveraging only core components (i.e., therapeutic topics, key questions, and optional advice) while allowing LLMs to support the dialogue generation and manage dialogue flow dynamically through therapeutic strategy prediction, inspired by prior CSCW research that uses therapeutic strategies to enhance human counseling effectiveness~\cite{Therapeutic_Skills_Affect_Clinical,MI_Strategies_cscw} as well as NLP studies~\cite{cos_xsun,cos} that instruct LLMs to follow evidence-based therapeutic principles.}
Results from Study 2 showed that LLM-SSAG performed comparably to LLM-SAG (Prompt) across most evaluation metrics, \minor{addressing our RQ2.}
This demonstrated that SSAG can maintain therapeutic structure by anchoring each response to a predicted MI strategy (e.g., asking a question, offering advice, or providing a reflection), ensuring alignment with evidence-based therapeutic behaviors while allowing flexible generation. 
\revise{By guiding LLMs through strategy prediction, especially reflective listening, SSAG tried to mitigate the over-reliance on therapeutic ``method'' (e.g., CBT) seen in prior work~\cite{therapy_cbt_llm} and supports more engaging, empathetic dialogues.}
This structured-yet-adaptive design mirrors techniques like Retrieval-Augmented Generation (RAG)~\cite{rag}, but with greater controllability for expert oversight.

\revise{The choice of MI and CBT balanced LLMs by combining MI’s emphasis on empathy and engagement with CBT’s structured, goal-oriented interventions.}
\minor{However, SSAG is not limited to these two focused frameworks. Its stepwise, strategy-level design allows integration of other therapeutic techniques by substituting proper expert-defined strategies.
Because SSAG operates at the level of therapeutic strategies rather than fixed language patterns, it is also adaptable to multilingual and culturally diverse contexts by accommodating different language norms and counseling styles.
}



\subsection{Implication: \minor{LLM-Supported Co-Development} of Psychotherapy Chatbots}




Psychotherapy chatbots, as socially impactful tools, hold promise for expanding access to mental health support, particularly in underserved communities. However, their development has traditionally relied on labor-intensive expert scripting, which limits scalability.
Our work addressed this challenge by first evaluating alignment approaches and then introducing Script-Strategy Aligned Generation (SSAG), a method designed to reduce dependence on fully \minor{expert-scripted} dialogues and enable more efficient co-development of scalable and explainable LLM-powered psychotherapy chatbots between AI (i.e., LLM), developers, and domain experts.

\revise{From a technical perspective,} Study 1 confirmed that prompting is a more scalable and resource-efficient alignment method than fine-tuning. Prompting preserved therapeutic adherence while allowing for flexible, human-like interaction, making it suitable for conversational therapies such as MI and CBT, where structured intervention must be delivered through adaptable dialogue. 
Building on this, SSAG introduced a two-stage strategy-level alignment method that relies on partial expert input (i.e., therapeutic topics, key questions, and general advice), while delegating dialogue generation and flow management to the LLM via predicted MI strategies. 
\revise{Results from Study 2 showed that SSAG matched the performance of fully aligned SAG chatbots across key therapeutic metrics while requiring less than 40\% of expert-authored dialogues, offering strong evidence of reduced reliance on expert-crafted dialogue content, answering our RQ3.}
\minor{While we did not directly measure reductions in expert workload by SSAG, its design eliminates the need for domain experts to manually construct full dialogue trees or write structured prompts. This suggests potential for easing expert involvement during the chatbot development process.} For instance, experts can focus on defining high-level strategies and therapeutical questions, while the LLM handles context-specific generation, reducing the need for technical prompt engineering.
\minor{We acknowledge that this only partially demonstrates collaborative development. While SSAG facilitates division of labor between domain experts and AI, the degree of collaboration was not empirically studied. 
Future CSCW research should more directly evaluate expert-AI workflows using participatory design methods, task-based workload assessments, or longitudinal co-development trials to rigorously assess reductions in human workload of both expert and developer.}

\revise{From a CSCW and human-AI collaboration perspective,} SSAG advances the vision of cooperative development in expert-driven and sensitive domains. 
\revise{SSAG positions LLMs as flexible yet controllable collaborators that support experts in constructing adaptive, therapeutically grounded dialogues, operating within the bounds of expert-guided therapeutic strategies.}
\revise{This draws on prior CSCW work on human-AI co-creation frameworks~\cite{Co-Creation-1} and domain-specific chatbot co-development~\cite{Plug_play}, demonstrating how structured human-AI collaboration can scale innovation in mental health technologies.}
\revise{Although we evaluated it in a semi-controlled setting, deployment in unsupervised settings still introduces ethical and safety concerns.
Real-world deployment requires additional safeguards, including human-in-the-loop oversight~\cite{humaninloop_1,humaninloop_2}, real-time validation mechanisms, and adherence to ethical standards. These measures are especially critical in unsupervised or high-risk settings like psychotherapy.}

\revise{Taken together, SSAG aligns with CSCW’s mission to foster computer-supported cooperative work in socially impactful domains.} By reducing development barriers and enabling scalable, explainable, and expert-aligned AI, SSAG opens pathways for delivering cost-effective and accessible mental health support at scale, particularly in underserved or resource-limited fields. 
\minor{Its design facilitates a scalable and efficient development between experts and developers, illustrating how AI can be harnessed in ways that are explainable, domain-aligned, and co-developed.}
Through this lens, this work serves as both a technical advancement and a CSCW-relevant contribution, representing a promising step toward building cooperative, expert-informed and socially impactful AI technologies that advance mental health support systems and the broader goals of CSCW.

%% file: 8-Conclusion.tex


                

\section{Limitations and Future Work}
Despite the contributions of this work, we acknowledge several limitations that remain for future exploration.

First, while the expert-crafted dialogue scripts used in this study are valuable, the reliance on such data introduces potential biases, which could impact the quality of LLM alignment. Its limited size and sole focus on MI and CBT, may not fully capture the diversity of real-world therapeutic interactions or generalize well to other more intensive forms of psychotherapy that require different conversational structures and techniques. 
Future work should expand datasets to cover a wider range of therapeutic scenarios enhancing SSAG's generalizability and robustness.
\minor{Moreover, although SSAG demonstrated reduced reliance on full scripting, requiring fewer than 40\% of expert-authored utterances compared to SAG, we did not directly measure expert workload or time savings. Thus, the efficiency claim should be interpreted as indirect. Future work should include task-based workload assessments or participatory design studies to evaluate the reduction of human workload more rigorously.} 

Second, although Study 2 included a 10-day field evaluation, the sample size was modest. Given the personalized nature of psychotherapy, user evaluations may vary widely and the reliance on self-reported feedback limits generalizability of our findings. Future work should involve more diverse samples and incorporate expert assessments for more robust evaluations.
\minor{While the study duration was relatively short, prior HCI and digital mental health research has employed comparable intervention periods (e.g., 2 weeks) to assess the effectiveness of such tools or applications~\cite{study_duration_1, study_duration_2, study_duration_3}.}
\minor{Nonetheless, human-perceived AI empathy and warmth may evolve over time. Longitudinal studies are needed to examine how these perceptions and therapeutic outcomes develop with continued use in real-world deployments.}

Third, despite efforts to ensure control and explainability, \revise{deploying LLM-powered chatbots in psychotherapy raises ongoing ethical and safety concerns}. Even when aligned with expert-crafted dialogues, LLMs may produce unpredictable or inappropriate responses that pose risks in sensitive mental health contexts. Addressing these concerns requires safeguards such as real-time monitoring, fallback mechanisms for high-risk situations, and stronger data privacy and explainability protocols.

\camera{Fourth, participants’ prior exposure to chatbots or similar AI-driven services may have influenced their perceptions and interactions. Those with more experience might have had different expectations or levels of comfort, potentially affecting their evaluations of the chatbots. 
While we did not explicitly control the prior exposure for participant recruitment, the counterbalanced study design helps mitigate systematic bias. Future work could explore this variable more directly with stratified analyses or larger samples to better understand how prior exposure shapes user perceptions in chatbots in psychotherapeutic contexts.}


While this work advances the alignment of LLMs with expert-crafted dialogues, addressing these limitations is essential to ensure the scalability, generalizability and ethical deployment of LLM-powered psychotherapy chatbots.


\section{Conclusion}

Our work provided initial evidence that aligning LLMs with expert-crafted dialogues may significantly enhance their performance and user perceptions in chatbot-delivered psychotherapy for behavioral interventions. By creating a dataset of expert-crafted dialogue scripts for Motivational Interviewing (MI) and Cognitive Behavioral Therapy (CBT), we conducted empirical studies comparing rule-based, pure LLM, and LLM-aligned chatbots to evaluate their conversational quality and therapeutic effectiveness. 
Our proposed approach, Script-Strategy Aligned Generation (SSAG), integrated human expertise with LLMs, resulting in controllable and engaging chatbots. 
This approach mitigated data scarcity, reduced reliance on expert input, and enhanced the controllability of LLMs in delivering psychotherapy for behavioral interventions. 
The findings underscored the importance of human-guided LLMs for controllable and explainable digital mental health tools and contributed to the field of CSCW by advancing scalable, cost-effective and LLM-supported development of psychotherapy chatbots, a promising path for future research in digital mental healthcare and health behavioral intervention.

%% file: appendix_dialogues.tex
\clearpage
\newpage
\onecolumn

\begin{appendices}
\renewcommand{\thesection}{Appendix A.1}
\section{Overview of the Expert-Crafted Dialogue Scripts}
\label{appendix_dialogue}
\end{appendices}

We provide a detailed overview of the dataset we created in this work and introduced in Section~\ref{dataset}.

\begin{table}[!h]
\centering
\scriptsize
\renewcommand{\arraystretch}{1.16}
\begin{tabularx}{\columnwidth}{
p{1.6cm} > 
{\raggedright\arraybackslash}p{4.0cm} >
{\raggedright\arraybackslash}p{5.2cm} >
{\raggedright\arraybackslash}X}
\toprule
\textbf{Technique} & \textbf{Psychotherapeutic Topic} & \textbf{Description of Each Topic} & \textbf{Num of Question / Advice / Total Utterance per Topic} \\
\midrule

\multirow{1}{*}{\textbf{Overall}}
 &  &  &  501 / 203 / 1876 \\ \hline

\multirow{22}{*}{MI} 
 & Greeting & Establishes rapport and initiates a dialogue. & 2 / 0 / 9\\ \cline{2-4}

 & Transition topic & Facilitates smooth shifts between therapeutic topics. &  8 / 0 / 18\\ \cline{2-4}

 & General introduction & Introduces the purpose of the conversation. & 14 / 11 / 54 \\ \cline{2-4}
 
 & Status of physical activity & Explores current levels of physical activity to understand baseline behavior. & 15 / 13 / 81 \\ \cline{2-4}
 
 & Handling Bad Weather & Discusses strategies to maintain motivation during poor weather. & 14 / 14 / 65  \\ \cline{2-4}
 
 & Sleep Quality & Gathers information about sleep habits and explores links to energy/motivation. & 3 / 12 / 32 \\ \cline{2-4}
 
 & Setting a Goal & Helps the user identify a meaningful, self-directed behavioral goal. & 8 / 4 / 36 \\ \cline{2-4}
 
 & Setting a Goal follow up & Revisits previously set goals to reflect on progress, challenges, or adjustments. & 27 / 2 / 118 \\ \cline{2-4}
 
 & Motivational onboarding for energy & Educates and motivates the user to consider movement as a way to boost energy. & 11 / 9 / 50 \\ \cline{2-4}
 
 & Rating confidence of physical activity & Uses confidence scales to explore and enhance self-efficacy for being active. & 10 / 11 / 68 \\ \cline{2-4}
 
 & Evoking self-efficacy & Encourages the user to recognize their own strengths and capacity for change. & 10 / 16 / 103 \\ \cline{2-4}
 
 & Improving motivation & Identifies personal reasons for change to boost intrinsic motivation. &  39 / 10 / 117 \\ \hline

\multirow{28}{*}{CBT Practices} 
 & Supportive social environment & Encourages reflection on positive social support for behavior change. & 15 / 7 / 101 \\ \cline{2-4}
 
 & Supportive social environment follow up & Assesses ongoing support and identifies ways to strengthen helpful connections. & 22 / 5 / 79 \\ \cline{2-4}
 
 & Mindfulness & Introduces present-moment awareness as a way to manage thoughts and emotions. & 17 / 5 / 52 \\ \cline{2-4}
 
 & Mindfulness follow up & Reviews practice experience and reinforces benefits of mindfulness. & 7 / 8 / 27 \\ \cline{2-4}
 
 & All or nothing thinking & Identifies black-and-white thinking patterns that can block progress. & 47 / 1 / 122 \\ \cline{2-4}
 
 & All or nothing thinking follow up & Applies alternative thinking strategies to reduce rigid thought patterns. & 22 / 2 / 63 \\ \cline{2-4}
 
 & Should-statements & Identifies unhelpful internal rules (e.g., “I should exercise every day”). & 31 / 5 / 73 \\ \cline{2-4}
 
 & Should-statements follow up & Challenges and reframes unrealistic expectations or self-criticism. & 15 / 1 / 49 \\ \cline{2-4}
 
 & Implement Intentions & Helps translate intentions into action using planning and cue-based strategies. & 45 / 11 / 130 \\ \cline{2-4}
 
 & Stress management & Teaches techniques (e.g., reframing, breathing) to cope with stress effectively. & 42 / 9 / 107 \\ \cline{2-4}
 
 & Habit building & Supports creation of consistent, repeatable behaviors that lead to long-term motivation and change. & 22 / 17 / 103 \\ \cline{2-4}
 
 & Time management & Offers tools to prioritize and schedule healthy activities realistically. & 18 / 12 / 78 \\ \cline{2-4}
 
 & Unsupportive social environment & Identifies social barriers and develops strategies to handle them constructively. & 25 / 13 / 91 \\ \cline{2-4}
 
 & Small wins & Encourages small, achievable actions to build momentum and confidence. & 12 / 5 / 50 \\ 
 

\bottomrule
\end{tabularx}
\vspace{1mm}
\caption{Overview of expert-crafted dialogue scripts.
"Num of Question / Advice / Total Utterance per Topic" indicates the number of therapeutic questions, general advice, and total utterances included in the dialogue script for each psychotherapeutic topic.}
\vspace{-3.6mm}
\end{table}


\clearpage

\begin{appendices}
\renewcommand{\thesection}{Appendix A.2}
\section{Example of Expert-Crafted Dialogue in Motivational Interviewing (MI) with the Topic "Physical Activity Level"}
\label{appendix_dialogue_mi}
\end{appendices}

\begin{figure*}[!ht]
\centering
\includegraphics[width=0.998\textwidth]{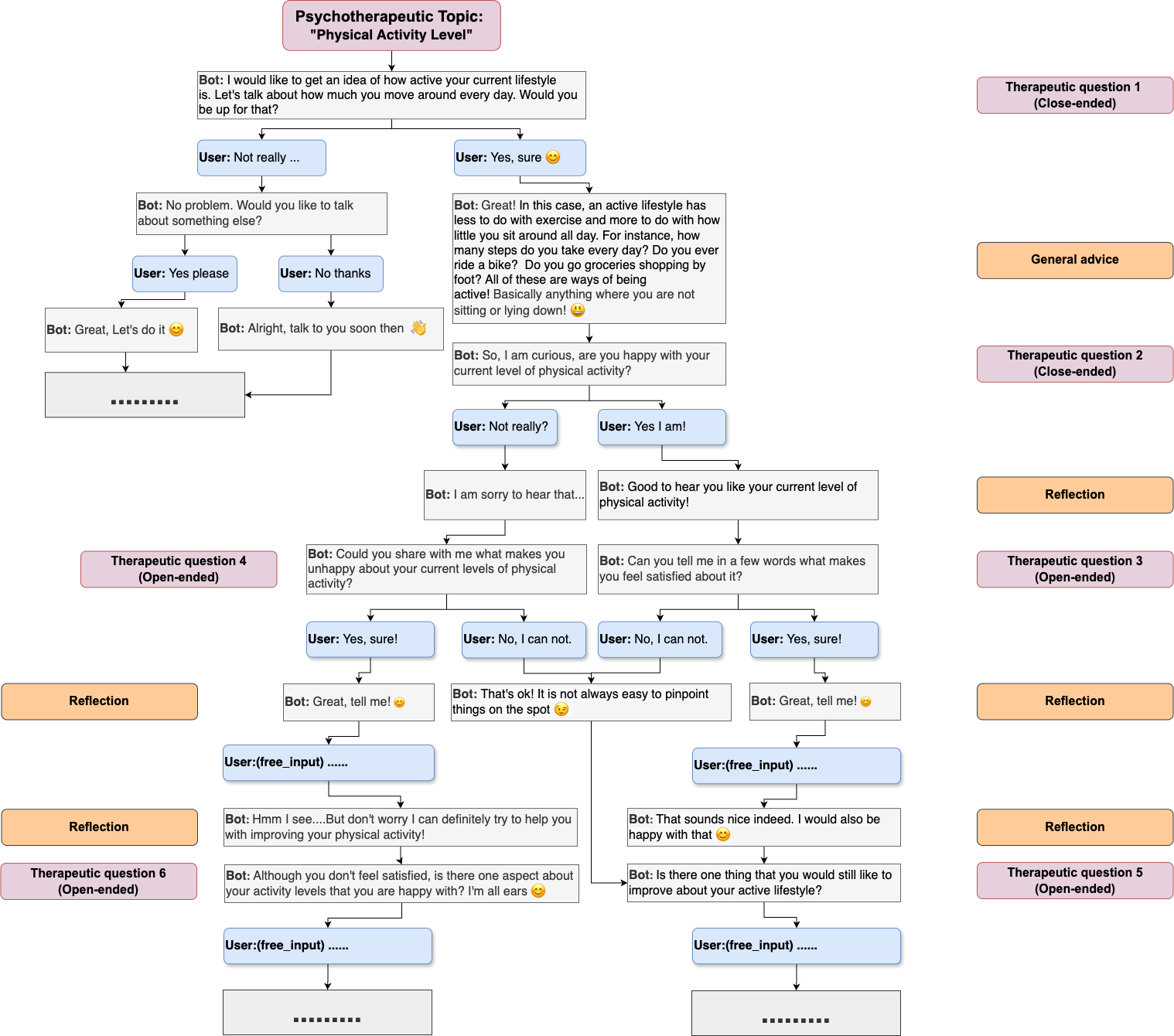}
\vspace{-0.0mm}
\caption{Examples of expert-crafted dialogues in motivational interviewing (MI) with psychotherapeutic topic "physical activity level". The expert-crafted dialogues contain 1) psychotherapeutic topics, 2) therapeutic questions (both open-/close-ended questions), 3) reflections, 4) general advice, 5) dialogue flows pre-designed by experts.}
\label{fig:dialogue_example_pa}
\vspace{-4.0mm}
\end{figure*}


\clearpage

\begin{appendices}
\renewcommand{\thesection}{Appendix A.3}
\section{Example of Expert-Crafted Dialogue in Cognitive Behavioral Therapy (CBT) with the Topic "Should Statement"}
\label{appendix_dialogue_cbt}
\end{appendices}

\begin{figure*}[!ht]
\centering
\includegraphics[width=0.998\textwidth]{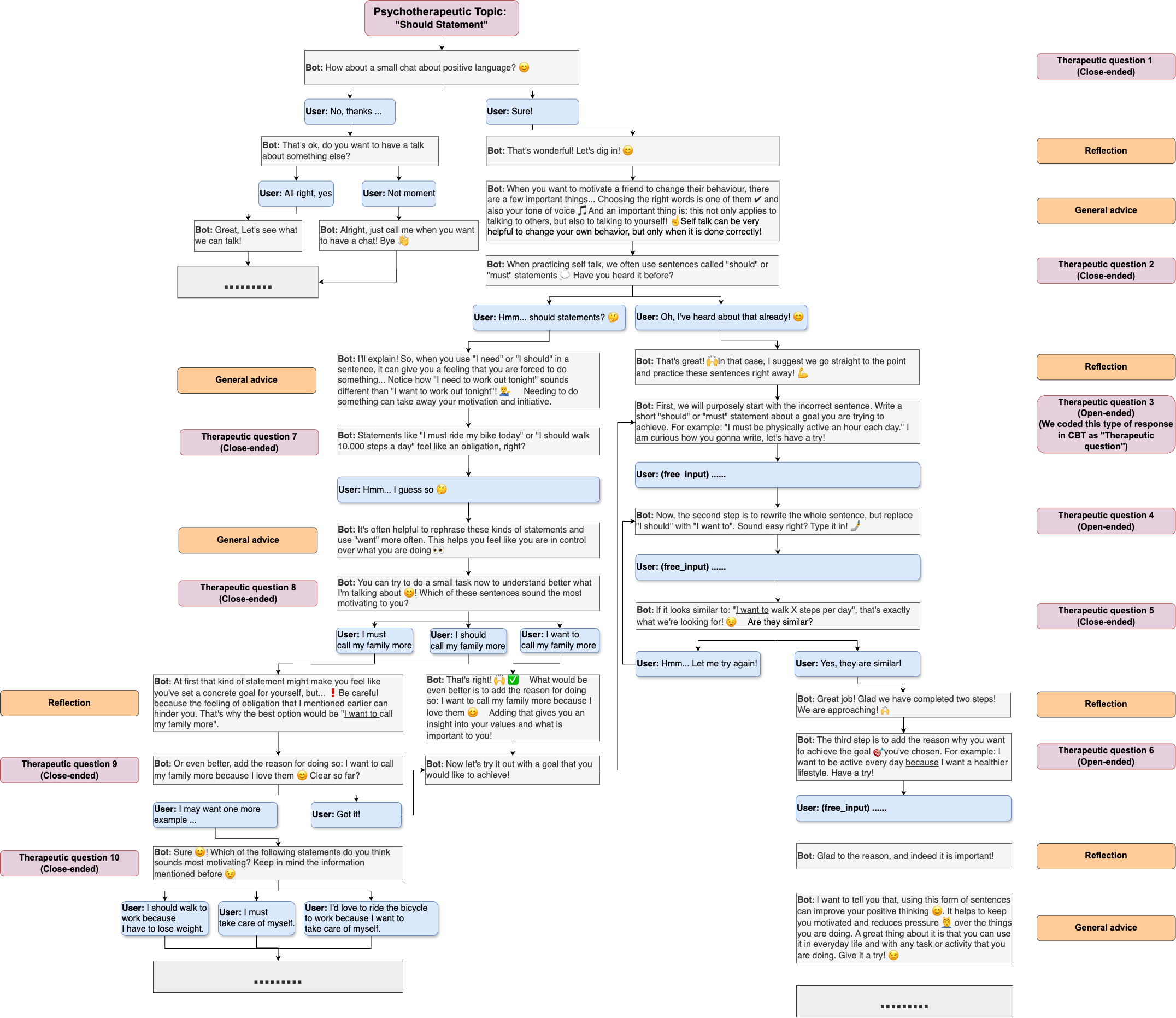}
\vspace{-0.0mm}
\caption{Examples of expert-crafted dialogues in cognitive behavioral therapy (CBT) with psychotherapeutic topic "should statement". The expert-crafted dialogues contain 1) psychotherapeutic topics, 2) therapeutic questions (both open-/close-ended questions), 3) reflections, 4) general advice, 5) dialogue flows pre-designed by experts.}
\label{fig:dialogue_example_pa}
\vspace{-4.0mm}
\end{figure*}

%% file: appendix_prompts.tex
\clearpage

\begin{appendices}
\renewcommand{\thesection}{Appendix B.1}
\section{Prompt Template Used in the Study: Pure LLM \& LLM-SAG (FT)}
\label{appendix:prompt-template-1}
\end{appendices}

\begin{table}[ht!]
\centering
\footnotesize
\renewcommand{\arraystretch}{1.4}
\begin{tabularx}{\textwidth}{>{\raggedright\arraybackslash}p{0.26\textwidth} >
{\raggedright\arraybackslash}X} 
\hline
\multicolumn{1}{l}{\textbf{Prompt Components}} & \multicolumn{1}{l}{Content} \\
\hline

\textbf{Conversation context} & 
\textbf{[The continuously accumulated context of the conversation]} \newline 

Conversation context: \newline 

Therapist: [...] \newline
Client: [...] \newline
Therapist: [...] \newline
Client: [...]  \newline
[......] \newline \\ 
\hline

\textbf{Task instruction} &
\textbf{[The base instructions to explain the generation task]} \newline

Task:\newline

You are a psychotherapist conducting a session to promote healthier behavior using Motivational Interviewing (or Cognitive Behavioral Therapy). The current therapeutic topic is [given topic]. \newline

The definition of this topic is [description of the given topic]. \newline \\ 
\hline

\end{tabularx}
\caption{The prompt template used in the Study 1 for pure LLM chatbots and LLM-SAG (FT) chatbots.}
\end{table}


\clearpage

\label{apndx:prompt-template-2}
\begin{appendices}
\renewcommand{\thesection}{Appendix B.2}
\section{Prompt Template Used in the Study: LLM-SAG (Prompt)}
\label{appendix:prompt-template-2}
\end{appendices}

\begin{table}[ht!]
\centering
\footnotesize
\renewcommand{\arraystretch}{1.4}
\begin{tabularx}{\textwidth}{>{\raggedright\arraybackslash}p{0.26\textwidth} >
{\raggedright\arraybackslash}X} 
\hline
\multicolumn{1}{l}{\textbf{Prompt Components}} & \multicolumn{1}{l}{Content} \\
\hline

\textbf{Conversation context} & 
\textbf{[The continuously accumulated context of the conversation]} \newline 

Conversation context: \newline 

Therapist: [...] \newline
Client: [...] \newline
Therapist: [...] \newline
Client: [...]  \newline
[......] \newline \\ 
\hline

\textbf{Expert-crafted dialogue script} & 
\textbf{[The expert-crafted dialogue script (including both dialogue content and flow) for the specific psychotherapeutic topic of the current session.]} \newline 

All dialogue flows in expert-crafted dialogue scripts of [given topic]: \newline 

Flow 1: [Pairs of Therapist-Client Dialogues] \newline
Flow 2: [Pairs of Therapist-Client Dialogues] \newline
Flow 3: [Pairs of Therapist-Client Dialogues] \newline
Flow 4: [Pairs of Therapist-Client Dialogues] \newline
[......] \newline \\ 
\hline

\textbf{Task instruction} &
\textbf{[The base instructions to explain the generation task]} \newline

Task:\newline

You are a psychotherapist conducting a session to promote healthier behavior using Motivational Interviewing (or Cognitive Behavioral Therapy) following tree-structured dialogue flows. The current therapeutic topic is [given topic]. \newline
The definition of this topic is [description of the given topic].\newline

Rules to follow:\newline
1. Identify the relevant dialogue tree based on the current conversation context and user input. \newline
2. At each conversational turn, please use the Tree-of-Thought technique about the decision point and use Breadth-First Search to go level-by-level in the dialogue tree. Do not explore deeper branches unless the current path is chosen.\newline
3. If no exact match is found, please intelligently adapt while staying close to the current dialogue flow/tree.\newline
4. You can give tailored responses to user input and provide dialogue diversity, but please keep staying strictly to the given dialogue flow/tree.\newline
5. Follow the dialogue tree structure and generate possible next responses based on relevance and coherence and with a clear format.\newline
6. If the ongoing conversational topic is finished, please say goodbye to the user and ask the user to go click the button on the left to evaluate this chatbot. \newline \\ 
\hline

\end{tabularx}
\caption{The prompt template used in Study 1 and 2 for the LLM-SAG (Prompt) chatbots.}
\end{table}


\clearpage

\begin{appendices}
\renewcommand{\thesection}{Appendix B.3}
\section{Prompt Template Used in the Study: LLM-SSAG}
\label{appendix:prompt-template-3}
\end{appendices}

\begin{table}[ht!]
\centering
\scriptsize
\renewcommand{\arraystretch}{1.36}
\begin{tabularx}{\textwidth}{>{\raggedright\arraybackslash}p{0.14\textwidth} >
{\raggedright\arraybackslash}p{0.20\textwidth} >
{\raggedright\arraybackslash}X} 
\hline
\multicolumn{1}{l}{\textbf{Step}} & \multicolumn{1}{l}{\textbf{Prompt Components}} & \multicolumn{1}{l}{Content} \\
\hline

\textbf{Step 1 of SSAG} & & \\  \hline

& \textbf{Conversation context} & 
\textbf{[The continuously accumulated context of the conversation]} \newline 

Conversation context: \newline 

Therapist: [...] \newline
Client: [...] \newline
Therapist: [...] \newline
Client: [...]  \newline
[......] \newline \\ 
\cline{2-3}

& \textbf{Therapeutic strategy} & 
\textbf{[The definition of the therapeutic strategy]} \newline 

The definition of the therapeutic strategy: \newline 

"reflection": [...] \newline 
"question": [...] \newline 
"advice": [...] \newline \\ 
\cline{2-3}

& \textbf{Task instruction} & 
\textbf{[The base instructions to explain the generation task]} \newline 

Task: \newline 

As a psychotherapist of Motivational Interviewing, please predict the next appropriate therapeutic strategy only from the set ["reflection", "question", "advice"] based on the conversation context.
The next therapeutic strategy is (are): 
\newline \\ 
\hline

\textbf{Step 2 of SSAG} & & \\  \hline

& \textbf{Conversation context} & 
\textbf{[The continuously accumulated context of the conversation]} \newline 

Conversation context: \newline 

Therapist: [...] \newline
Client: [...] \newline
Therapist: [...] \newline
Client: [...]  \newline
[......] \newline \\ 
\cline{2-3}

& \textbf{Next therapeutic strategy} & 
\textbf{[The next therapeutic strategy of the chatbot response]} \newline

The next therapeutic strategy is (or strategies are): \newline 
``Reflection'' or ``Question'' or ``Advice'' or their combination based on the prediction from Step 1. \newline 

The definition of the therapeutic strategy: [definition of the predicted therapeutic strategy]. \newline \\

\cline{2-3}

& \textbf{Task instruction} &
\textbf{[The base instructions to explain the generation task]} \newline

Task:\newline

You are a psychotherapist conducting a session to promote healthier behavior using Motivational Interviewing (or Cognitive Behavioral Therapy). The current therapeutic topic is [given topic]. \newline

The definition of this topic is [description of the given topic]. \newline 

You should generate the response strictly aligned with the next therapeutic strategy above.\newline \\ 
\hline

\end{tabularx}
\caption{The prompt template used in Study 2 for the LLM-SSAG chatbots.}
\end{table}

%% file: appendix_implement.tex
\clearpage

\begin{appendices}
\renewcommand{\thesection}{Appendix C}
\section{Implementation Details of Generations with LLMs}
\label{appendix:implement_details}
\end{appendices}

We utilized the August 2024 edition of GPT-4o, coded as \texttt{gpt-4o-2024-08-06}~\footnote{\url{https://platform.openai.com/docs/models/gpt-4o}\label{ft:gpt4}}.
We used \texttt{openai} Python library to generate with GPT-4. 
We opted for default hyperparameters, including the temperature as default to control the randomness of generation. 
The models were used in compliance with their respective licenses and terms at the time of the study. OpenAI provides a Terms of Use\footnote{\url{https://openai.com/policies/terms-of-use}}.


